\documentclass[12pt]{JHEP3}

\title{Natural Priors, CMSSM Fits and LHC Weather Forecasts}

\author{Benjamin C Allanach$^{1}$, Kyle Cranmer$^2$, Christopher G Lester$^{3}$ and Arne M
  Weber$^{4}$ \\ 
$^{1}$ DAMTP, CMS, Wilberforce Road, Cambridge CB3 0WA, UK\\
$^{1}$ Dept. of Physics, New York University, New York, USA\\
$^{3}$ Cavendish Laboratory, J.J. Thomson Avenue, Cambridge CB3 0HE, UK\\
$^4$ Max Planck Inst.\ f\"{u}r Phys., F\"{o}hringer Ring 6, D-80805 Munich,
  Germany\\ 
}

\abstract{Previous LHC forecasts for the constrained minimal
supersymmetric standard model (CMSSM), based on
current astrophysical and laboratory measurements, have used priors
that are flat in the 
parameter $\tan \beta$, while being constrained to postdict the central
experimental value of $M_Z$.  
We construct a different,
new and more natural prior with a measure
in $\mu$ and $B$ (the more
fundamental MSSM parameters from which $\tan \beta$ and $M_Z$ are
actually derived).  We find that as a consequence this choice leads to
a well defined fine-tuning measure in the parameter space. We
investigate the effect of such on global CMSSM fits to indirect
constraints, providing posterior probability distributions for Large
Hadron Collider (LHC) sparticle production cross sections. The change
in priors has a significant effect, strongly suppressing the
pseudoscalar Higgs boson dark matter annihilation region, and
diminishing the probable values of sparticle masses. We also show how to
interpret fit information from a Markov Chain Monte Carlo in a frequentist 
fashion; namely by using the profile likelihood. Bayesian and frequentist
interpretations of CMSSM fits are compared and contrasted.
}

 \newlength{\wth}
\usepackage[dvips]{epsfig}

\newcommand{\fourgraphs}[4]{%
 \unitlength=1.1in
 \begin{picture}(5.8,4.4)(0.3,0.3)
\put(0,2.4){\put(0.5,0){\epsfig{file=#1, width=0.698 \wth}}
 \put(2.9,0){\epsfig{file=#2, width=0.698 \wth}}
 \put(0.5,2.2){(a)}
 \put(3,2.2){(b)}}
\put(0,0){\put(0.5,0){\epsfig{file=#3, width=0.698 \wth}}
 \put(2.9,0){\epsfig{file=#4, width=0.698 \wth}}
 \put(0.5,2.2){(c)}
 \put(3,2.2){(d)}}
 \end{picture}
}
\newcommand{\sixgraphs}[6]{%
\unitlength=0.9in
\begin{picture}(5,7.5)
\put(0,5){\epsfig{file=#1.eps, width=2.3in}}
\put(2.5,5){\epsfig{file=#2.eps, width=2.3in}}
\put(0,0){\epsfig{file=#5.eps, width=2.3in}}
\put(2.5,0){\epsfig{file=#6.eps, width=2.3in}}
\put(0,2.5){\epsfig{file=#3.eps, width=2.3in}}
\put(2.5,2.5){\epsfig{file=#4.eps, width=2.3in}}
\put(0,4.7){(c)}
\put(0,7.2){(a)}
\put(2.5,7.2){(b)}
\put(0,2.2){(e)}
\put(2.5,2.2){(f)}
\put(2.5,4.7){(d)}
\end{picture}}

\newcommand{\fourgraphst}[4]{%
 \unitlength=1.1in
 \begin{picture}(5.8,4)(0.5,0.4)
\put(0,2){\put(-0.04,2.54){\epsfig{file=#1, width=0.698 \wth,angle=270}}
 \put(0.85,0.5){\epsfig{file=#12, width=0.68 \wth}}
 \put(2.66,2.54){\epsfig{file=#2, width=0.698 \wth, angle=270}}
 \put(3.55,0.5){\epsfig{file=#22, width=0.68 \wth}}
 \put(0.5,2.1){(a)}
 \put(3.2,2.1){(b)}
}
\put(0,0){\put(-0.04,2.54){\epsfig{file=#3, width=0.698 \wth,angle=270}}
 \put(0.85,0.5){\epsfig{file=#32, width=0.68 \wth}}
 \put(2.66,2.54){\epsfig{file=#4, width=0.698 \wth, angle=270}}
 \put(3.55,0.5){\epsfig{file=#42, width=0.68 \wth}}
 \put(0.5,2.1){(c)}
 \put(3.2,2.1){(d)}
}
 \end{picture}
}

\newcommand{\fourgraphstt}[4]{%
 \unitlength=1.1in
 \begin{picture}(5.8,4)(0.5,0.4)
\put(0,2){\put(0.45,0.15){\epsfig{file=#1, width=0.6 \wth}}
  \put(2.66,2.54){\epsfig{file=#2, width=0.698 \wth, angle=270}}
  \put(3.55,0.5){\epsfig{file=#22, width=0.68 \wth}}
  \put(0.5,2.1){(a)}
  \put(3.2,2.1){(b)}
}
\put(0,0){\put(-0.04,2.54){\epsfig{file=#3, width=0.698 \wth,angle=270}}
 \put(0.85,0.5){\epsfig{file=#32, width=0.68 \wth}}
 \put(2.66,2.54){\epsfig{file=#4, width=0.698 \wth, angle=270}}
 \put(3.55,0.5){\epsfig{file=#42, width=0.68 \wth}}
 \put(0.5,2.1){(c)}
 \put(3.2,2.1){(d)}
}
 \end{picture}
}

\newcommand{\twographst}[2]{%
 \unitlength=1.1in
 \begin{picture}(5.8,2.3)(0.5,0.25)
 \put(0.7,0.3){\epsfig{file=#1, width=0.6 \wth}}
 \put(2.66,2.54){\epsfig{file=#2, width=0.698 \wth, angle=270}}
 \put(3.56,0.5){\epsfig{file=#22, width=0.68 \wth}}
 \put(0.5,2.1){(a)}
 \put(3.2,2.1){(b)}
 \end{picture}
}
\newcommand{\twographs}[2]{%
 \unitlength=1.1in
 \begin{picture}(5.8,2.3)(0.5,0.25)
 \put(-0.04,2.54){\epsfig{file=#1, width=0.698 \wth,angle=270}}
 \put(0.85,0.5){\epsfig{file=#12, width=0.68 \wth}}
 \put(2.66,2.54){\epsfig{file=#2, width=0.698 \wth, angle=270}}
 \put(3.56,0.5){\epsfig{file=#22, width=0.68 \wth}}
 \put(0.5,2.1){(a)}
 \put(3.2,2.1){(b)}
 \end{picture}
}

\newcommand{\eightgraphs}[8]{%
 \unitlength=1in
 \begin{picture}(6,6)(0,0)
\put(0,4){\epsfig{file=#1, width=2 in}}
\put(2,4){\epsfig{file=#2, width=2 in}}
\put(4,4){\epsfig{file=#3, width=2 in}}
\put(0,2){\epsfig{file=#4, width=2 in}}
\put(2,2){\epsfig{file=#5, width=2 in}}
\put(4,2){\epsfig{file=#6, width=2 in}}
\put(1,0){\epsfig{file=#7, width=2 in}}
\put(3,0){\epsfig{file=#82, width=2 in}}
\put(3.8,0.6){\epsfig{file=#81, width=1 in}}
\put(0,5.8){(a)}
\put(2,5.8){(b)}
\put(4,5.8){(c)}
\put(0,3.8){(d)}
\put(2,3.8){(e)}
\put(4,3.8){(f)}
\put(1,1.8){(g)}
\put(3,1.8){(h)}
 \end{picture}
}

\preprint{DAMTP-2007-18\\ Cavendish-HEP-2007-03 \\ MPP-2007-36} 

\begin{document}

\section{Introduction}

The impending start-up of the LHC makes this a potentially exciting
time for supersymmetric (SUSY) phenomenology.  Anticipating the
arrival of LHC data, a small industry has grown up aiming to forecast
the LHC's likely discoveries.  There are big differences between
nature of the questions answered by a forecast, and the questions that
will be answered by the experiments themselves when they have acquired
compelling data.  A weather forecast predicting ``severe rain in
Cambridgeshire at the end of the week'' should not be confused with a
discovery of water.  However, the forecast {\em is} \/something which
influences short-term flood plans and will set priorities within the
list of ``urgent repairs needed by flood defences''.

LHC weather forecasts for sparticle masses or cross sections set
priorities among signals needing to be investigated, or among
expensive Monte Carlo background samples competing to be generated.
Forecasts can influence the design parameters of future experiments
and colliders.  In advance of LHC we would like to have some sort of
idea of what luminosity will be required in order to detect and/or
measure supersymmetry. There is also the question of which signatures
are likely to be present.

In order to answer questions such as these, a programme of fits to
simple SUSY models has proceeded in the
literature~\cite{Ellis:2003si,Profumo:2004at,Baltz:2004aw,Ellis:2004tc,Stark:2005mp}.
The fits that we are interested in have made the universality
assumption on soft SUSY breaking parameters: the scalar masses are set
to be equal to $m_0$, the trilinear scalar couplings are set to be
$A_0$ multiplied by the corresponding Yukawa couplings and all gaugino
masses are set to be equal to $M_{1/2}$. Such assumptions, when
applied to the MSSM, are typically called mSUGRA or the constrained
minimal supersymmetric standard model.  The universality conditions
are typically imposed at a gauge unification scale $M_{GUT} \sim 2
\times 10^{16}$ GeV. The universality conditions are quite strong, but
allow phenomenological analysis of a varied subset of MSSM models. The
universality assumption is not unmotivated since, for example, several
string models~\cite{unistringmodels} predict MSSM universality.

Until recently, CMSSM fits have relied upon fixed input
parameters~\cite{dickpran,Robros,Ros2,Ellis:2003si,Profumo:2004at,Baltz:2004aw,Ellis:2004tc} 
in order to reduce the dimensionality of the CMSSM parameter space,
rendering scans viable. Such analyses provide a good idea of what are
the relevant physical processes in the various parts of parameter
space.  More recently, however, it has been realised that
many-parameter scans are feasible if one utilises a Markov Chain Monte
Carlo (MCMC)~\cite{Baltz:2004aw}.  Such scans were used to perform
multi-dimensional a Bayesian analysis of indirect
constraints~\cite{Allanach:2005kz}. A particularly important
constraint came from the relic density of dark matter $\Omega_{DM}
h^2$, assumed to consist solely of neutralinos, the lightest of which
is the lightest supersymmetric particle (LSP). Under the assumption of
a discrete symmetry such as $R-$parity, the LSP is stable and thus
still present in the universe after being thermally produced in the
big bang.  The results of ref.~\cite{Allanach:2005kz} were confirmed
by an independent study~\cite{deAustri:2006pe}, which also examined
the prospects of direct dark matter detection. Since then, a study of
the $\mu<0$ branch of the CMSSM was performed~\cite{darkSide} and
implications for Tevatron Higgs searches have been
discussed~\cite{rosz2}.

It is inevitable that LHC forecasts will contain a large degree of
uncertainty.  This is unavoidable as, in the absence of LHC data,
constraints are at best indirect and also few in number.  Within a
Bayesian framework, the components of the answer that are
incontestable
lie within a simple ``likelihood'' function, whereas the
parts which parameterise our ignorance concerning the nature of the
parameter space we are about to explore are rolled up into a prior.
By separating components into these two domains, we have an efficient
means of testing not only what the data is telling is about new
physics, but also of warning us of the degree to which the data is (or
isn't) compelling enough to disabuse us of any prior expectations we
may hold.

In \cite{Allanach:2005kz,deAustri:2006pe}, Bayesian statements were
made about the posterior probability density of the CMSSM, after indirect
data had been taken into account.  The final result of a Bayesian analysis
is the posterior probability density function (pdf), which in previous MCMC
fits, was set to be 
\begin{equation}
p(m_0, M_{1/2}, A_0, \tan \beta, s | \mbox{data}) = 
p(\mbox{data} | m_0, M_{1/2}, A_0, \tan \beta, s) 
\frac{p(m_0, M_{1/2}, A_0, \tan \beta, s)}{p(\mbox{data})} \label{bayes}
\end{equation}
for certain Standard Model (SM) inputs $s$ and ratio of the two MSSM
 Higgs vacuum expectation values $\tan \beta = v_2/v_1$. The
 likelihood $p(\mbox{data} | m_0, M_{1/2}, A_0, \tan \beta, s)$ is
 proportional to $e^{-\chi^2/2}$, where $\chi^2$ is the common
 statistical measure of disagreement between theoretical prediction
 and empirical measurement.  The prior 
$p(m_0,M_{1/2},$ $A_0,\tan\beta,s)$ 
was taken somewhat arbitrarily to be flat (i.e.\ equal to a
 constant) within some ranges of the parameters, and zero outside
 those ranges. Eq.~\ref{bayes} has an implied measure for the input
 parameter. If, for example, we wish to extract the posterior pdf for
 $m_0$, all other parameters are marginalised over
\begin{equation}
  p(m_0| \mbox{data}) = \int d M_{1/2}\ d A_0\ d\tan \beta \ d s \
p(m_0, M_{1/2},A_0, \tan \beta, s | \mbox{data}).
\end{equation}
Thus a flat prior in, say, $\tan \beta$ also corresponds to a choice
of measure in the marginalisation procedure: $\int d \tan \beta$.
Before one has a variety of accurate direct data (coming, for
instance, from the LHC), the results depend somewhat upon what prior
pdf is assumed. 

In all of the previous MCMC fits, Higgs potential parameters 
$\mu$ and $B$ were traded for $M_Z$ and $\tan \beta$ 
using the electroweak symmetry breaking conditions, 
which are obtained by minimising the MSSM
Higgs potential and obtaining the relations~\cite{BPMZ}:
\begin{eqnarray}
  \mu B &=& \frac{\sin 2 \beta}{2} ( {\bar m}_{H_1}^2 + {\bar m}_{H_2}^2 + 2
  \mu^2 ), 
  \label{muB} \\
  \mu^2 &=& \frac{{\bar m}_{H_1}^2 - {\bar m}_{H_2}^2 \tan^2 \beta}{\tan^2
  \beta -1} -   \frac{M_Z^2}{2}. \label{musq}
\end{eqnarray}
Eqs.~\ref{muB},\ref{musq} were applied at a scale 
$Q=\sqrt{m_{{\tilde t}_1}
      m_{{\tilde  t}_2}}$, i.e.\ the geometrical average of the two stop
masses\footnote{Higgs potential loop corrections are taken into 
      account by writing~\cite{BPMZ} ${\bar m}_{H_i} \equiv m_{H_i}^2 -
      t_i / v_i$, $t_i$ being the tadpoles of Higgs $i$ and $v_i$ being its
      vacuum expectation value.}. 
$|\mu|$ was set in order to obtain the
empirically measured central value of $M_Z$ in Eq.~\ref{musq} and then
Eq.~\ref{muB} was solved for $B$ for a given input value of $\tan \beta$ 
and sign$(\mu)$. 
The flat prior in $\tan \beta$ in Eq.~\ref{bayes} does not reflect the fact
that $\tan \beta$ (as well as $M_Z$) is a derived quantity from the more
fundamental parameters $\mu$, $B$. 
It also does not contain information about
regions of fine-tuned parameter space, which we may consider to be less likely
than regions which are less fine-tuned. 
Ref.~\cite{Giudice:2006sn} clearly illustrates
that if one includes $\mu$ as a fundamental MSSM parameter, LEP has ruled out
the majority of the natural region of MSSM parameter space. 

A conventional measure of fine-tuning~\cite{finetuning} is
\begin{equation}
f=\mbox{max}_p \left[\frac{d \ln M_Z^2}{d \ln p}\right], \label{ft}
\end{equation}
where the maximisation is over $p\in \{m_0, M_{1/2}, A_0, \mu,
B\}$. Here, Eq.~\ref{musq} is viewed as providing a prediction for $M_Z$ given
the other MSSM parameters. When the SUSY parameters are large, a cancellation
between various terms in Eq.~\ref{musq} must be present in order to give $M_Z$
at the experimentally measured value.
Eq.~\ref{ft} is supposed to provide a measure of how sensitive this
cancellation is to the initial parameters. In Ref.~\cite{Allanach:2006jc}, a 
prior $\propto 1/f$ was shown to produce fits that were not wildly different
to those with a 
flat prior, but the discrepancy illustrated the level of uncertainty
in the fits.  The new (arguably less arbitrary) prior discussed in
section~\ref{sec:REWSBpriors} will be seen to lead to much larger
differences.

Here, we extend the existing literature in two main ways: firstly, we
construct a natural prior in the more fundamental parameters $\mu$,
$B$, showing in passing that it can be seen to act as a check on
fine-tuning. We display the MCMC fit results from such priors. 
Secondly, we present posterior pdfs for LHC
supersymmetric (SUSY) production cross-sections. These have not been
calculated before. 
We also present a comparison with a more frequentist statistics oriented fit,
utilising the profile likelihood. The difference between the flat-priors
Bayesian analysis and the profile likelihood contains information about volume
effects in the marginalised dimensions of parameter space. We describe an
extremely simple and effective way to extract profile likelihood information
from the MCMC chains already obtained from the Bayesian analysis with flat
priors.  

In the proceeding section~\ref{sec:REWSBpriors}, we derive the new
more natural form for the prior distributions mentioned above. In
section~\ref{sec:likelihood}, we describe our calculation of the
likelihood. In section~\ref{sec:fits}, we investigate the
limits on parameter space and pdfs for sparticle masses resulting from 
the new more natural priors.  
We go on to
discuss what this prior-dependence means in terms of the ``baseline
SUSY production'' for 
the LHC, and find out what it tells us about the ``error-bars'' which
should be attached to this and earlier LHC forecasts.  
In section~\ref{sec:profile}, we present our results in the profile likelihood
format.
In the following
section~\ref{sec:crossec} we present pdfs for total SUSY production
cross-sections at the LHC\@. Section~\ref{sec:conc} contains a summary
and conclusions.  In Appendix~\ref{app:comp}, we compare the fit
results assuming the flat $\tan \beta$ priors with a well-known result
in the literature in order to find the cause of an apparent
discrepancy.

\section{Prior Distributions \label{sec:REWSBpriors}}

We wish to start with a measure defined in terms of fundamental parameters 
$\mu$ and $B$, hence
\begin{eqnarray}
p(\mbox{all data})&=& \int d \mu \ d B \ d  A_0 \ d m_0 \ d M_{1/2} \ ds\left[ 
p(m_0,M_{1/2},A_0,\mu,B,s)\right. \nonumber \\ 
&& \left. p(\mbox{all data} | m_0,M_{1/2},A_0,\mu,B,s) \right], \label{initial}
\end{eqnarray}
where $p(\mbox{all data} | m_0,M_{1/2},A_0,\mu,B,s)$ is the likelihood
of the data with respect to the CMSSM and $p(m_0,M_{1/2},A_0,\mu,B,s)$ is
the prior probability distribution for CMSSM and SM parameters.
Of these two terms, the former is well defined, while the latter is
open to a degree of interpretation due to the lack of pre-existing
constraints on $m_0$, $M_{1/2}$, $A_0$, $\mu$, and $B$\footnote{If an earlier
  experiment had already set clear 
constraints on $m_0$, $M_{1/2}$, $A_0$, $\mu$, $B$, then even
the prior would be well defined, being the result of that previous
experiment.  As things stand, however, we don't know anything about
the likely values of these parameters, and so the prior must encode
our ignorance/prejudice as best we can.}.  We may approximately
factorise the unambiguous likelihood into two independent pieces: one
for $M_Z$ and one for other data not including $M_Z$, the latter defined to be
$p(\mbox{data} | m_0,M_{1/2},A_0,\mu,B,s)$
\begin{eqnarray}
&&p(\mbox{all data} | m_0,M_{1/2},A_0,\mu,B,s)\nonumber\\
& \approx & p(\mbox{data} | m_0,M_{1/2},A_0,\mu,B,s) \times \nonumber 
p(M_Z | m_0,M_{1/2},A_0,\mu,B,s) \nonumber \\ 
& \approx & p(\mbox{data} | m_0,M_{1/2},A_0,\mu,B,s)\times \delta(M_Z - M_Z^{cen}).
\label{mzlike}
\end{eqnarray}
In the last step we have approximated the $M_Z$ likelihood by a delta
function on the central empirical value $M_Z^{cen}$ because its experimental
uncertainties are so tiny. According to the Particle Data Group~\cite{pdg},
the current world average measurement is $M_Z=91.1876 \pm 0.0021$ GeV.  

Using Eqs.~\ref{muB},\ref{musq} to calculate a Jacobian factor and substituting
Eq.~\ref{mzlike} into Eq.~\ref{initial}, we obtain
\begin{eqnarray}
p(\mbox{all data}) &\approx& 
\int d \tan \beta \ d  A_0 \ d m_0 \ d M_{1/2} \left[ 
 r(B, \mu, \tan \beta) \right.
\nonumber \\
&&\left. 
p(\mbox{data}|m_0,M_{1/2},A_0,\mu,B,s) p(m_0,M_{1/2},A_0,\mu,B,s) 
\right]_{M_Z=M_Z^{cen}},
\end{eqnarray}
where the condition $M_Z=M_Z^{cen}$ can be applied by using the constraints of
Eqs.~\ref{muB},\ref{musq} with $M_Z=M_Z^{cen}$. The Jacobian
factor 
\begin{equation}
r(B, \mu, \tan \beta) =
M_Z \left|
\frac{B }{\mu \tan \beta } \frac{\tan^2\beta - 1}{\tan^2 \beta + 1}
\right| \label{rewsb}
\end{equation} 
disfavours high values of $\tan \beta$ and $\mu/B$ 
and comes from our more natural initial parameterisation of the Higgs potential
parameters in terms of $\mu$, $B$.   We will refer below to $r(B,\mu,\tan
\beta)$ in Eq.~\ref{summary} as the ``REWSB 
  prior''. Note that, if we consider $B \rightarrow {\tilde B}\equiv \mu B$ to
  be more 
  fundamental than the parameter $B$, one loses the factor of $\mu$ in the
  denominator of $r$ and 
  by sending $\int d B \ d \mu \rightarrow \int d {\tilde
    B} \ d 
  \mu \ \mu$. However, in the present paper we retain $B$ as a fundamental
  parameter    because of its appearance in 
  many supergravity mediation models of SUSY breaking. 

It remains for us to define the prior, $p(m_0,M_{1/2},A_0,\mu,B,s)$, a
measure on the parameter space.  In our case, this prior must
represent our degree of belief in each part of the space, in advance
of the arrival of {\em any} \/experimental data.
There is no single ``right'' way of representing ignorance in a
prior\footnote{There are however plenty of ``wrong'' ways of
representing ignorance.  Choosing $p(m_0,M_{1/2},A_0,\mu,B,s) \propto
\delta(m_0-40 {\rm\ GeV}) {\left( \arctan{(A_0/B)} \right)} ^ {100}$
would clearly impose arbitrary and unjustifiable constraints on at
least three of the parameters!}, and so some subjectivity must enter
into our choice.  We must do our best to ensure that our prior is as
``even handed'' as possible.  It must give approximately equal
measures to regions of parameter space which seem equally plausible.
``Even handed'' need not mean ``flat'' however.  A prior flat in $m_0$
is not flat in $m_0^2$ and very non-flat in $\log {m_0}$.  We must do
our best to identify the important (and unimportant) characteristics of
each parameter.  If the absolute value of a parameter $m$ matters,
then flatness in $m$ may be appropriate.  If dynamic range in $m$ is
more expressive, then flatness in $1/m$ (giving equal weights to each
order of magnitude increase in $m$) may make sense. If only the size
of $m$ relative to some related scale $M$ is of importance, then a
prior concentrated near the origin in $\log(m/M)$ space may be more
appropriate.  The freedoms contained within these, to some degree
subjective, choices permit others to generate priors different from
our own, and thereby test the degree to which the data or the analysis
is compelling.  If the final results are sensitive to changes of
prior, then more data or a better analysis may be called for.

The core idea that we have chosen to encode in (and which therefore
defines) our prior on $m_0$, $M_{1/2}$, $A_0$, $\mu$, $B$, and $s$ may
be summarised as follows.  
(1) We define regions of parameter space
where there parameters all have similar orders of magnitude to be more
natural than those where they are vastly different.  For example we
regard $m_0 = 10^1$ eV, $M_{1/2} = 10^{20}$ eV as unnatural.  In effect,
we will use the distance measure between each parameter and a joint
`supersymmetry scale'' $M_S$ to define our prior. 
(2) We do not wish
to impose unity of scales at anything stronger than the order of
magnitude level.  
(3) We do not wish to presuppose any particular
scale for $M_S$ itself -- that is for the data to decide.

Putting these three principles together, we first define a measure
that would seem reasonable {\em were the supersymmetry scale of $M_S$
to be known}.  Later we will integrate out this dependence on
$M_S$. To begin with we factorise the prior probability density for a
given SUSY breaking scale $M_S$:
\begin{eqnarray}
p(m_0,M_{1/2},A_0,\mu,B,s | M_S) &=&p(m_0|M_S)\ p(M_{1/2}|M_S) \ 
p(A_0|M_S) \label{pfacs}
\\&& p(\mu |M_S) \ p(B|M_S) \ p(s)\nonumber,  
\end{eqnarray}
where we have assumed that the SM experimental inputs do not depend
upon $M_S$. This factorisation of priors could be changed to specialise for
particular models of SUSY breaking. For example, dilaton domination in
heterotic string models predicts $m_0=M_{1/2}=-A_0/\sqrt{3}$. In that case,
one would neglect the separate prior factors for $A_0$, $M_{1/2}$ and $m_0$ in
Eq.~\ref{pfacs}, leaving only one of them.
Since it is our intention to impose unity between $m_0$,
$M_{1/2}$, $A_0$ and $M_S$ at the ``order of magnitude'' level, we
take a prior probability density
\begin{equation}
p(m_0|M_S) = \frac{1}{\sqrt{2 \pi w^2} m_0} \exp \left(-
\frac{1}{2 w^2} \log^2
(\frac{m_0}{M_S}) \right). \label{priorM0}
\end{equation}
The normalising factor in front of the exponential ensures that
$\int_0^\infty d m_0 \ p(m_0 | M_S) = 1$. $w$ specifies the width of the
logarithmic exponential,
Eq.~\ref{priorM0} implies that $m_0$
is within a factor $e^w$ of $M_S$ at the ``1$\sigma$ level''
(i.e.\ with probability 68$\%$). 
We take analogous forms for $p(M_{1/2}|M_S)$ and $p(\mu \  | M_S)$, by
replacing $m_0$ in Eq.~\ref{priorM0} with $M_{1/2}$ and $|\mu|$
respectively.
Note in particular that our prior $p(\mu|M_S)$ favours  
superpotential parameter $\mu$ to be within an order of magnitude of  
$M_S$ and thus also within an order of magnitude of the soft breaking
parameters. This should be required by whichever model is responsible for 
solving the $\mu$ problem of the MSSM, for example the Giudice-Masiero
mechanism~\cite{gm}. 
 $A_0$ and $B$ are allowed to have positive or negative signs 
and values may pass through zero, 
so we chose a different form to Eq.~\ref{priorM0}
for their prior. However, we still expect that their order of
magnitude isn't much greater than $M_S$ and the prior probability density
\begin{equation}
p(A_0|M_S) = \frac{1}{\sqrt{2 \pi e^{2w}} M_S} \exp \left(-
\frac{1}{2 (e^{2w})} 
\frac{A_0^2}{M_S^2} \right), \label{priorA0}
\end{equation}
ensures that $|A_0|<e^w M_S$ at the 1$\sigma$ level. The prior probability
density of $B$ is given by 
Eq.~\ref{priorA0} with $A_0 \rightarrow  B$. 
We don't know $M_{S}$ a priori, so we marginalise over it:
\begin{eqnarray}
&&p(m_0,M_{1/2},A_0,\mu,B)
=\int_0^{\infty} d M_S\ p(m_0,M_{1/2},A_0,\mu,B | M_S)\ p(M_S)
\label{priorsummary} \\
&=&\frac{1}{ (2 \pi)^{5/2} w^5 m_0 |\mu| M_{1/2}}\int_0^{\infty} \frac{d
  M_{S}}{M_S^2} 
\exp \left[-\frac{1}{2 w^2} \left( 
\log^2 (\frac{m_0}{M_S}) +
\log^2 (\frac{|\mu|}{M_S}) + 
 \right. \right. \nonumber \\
&& \left. \left. 
\log^2 (\frac{M_{1/2}}{M_S}) + 
\frac{w^2 A_0^2}{e^{2w} M_{S}^2} + \frac{w^2 B^2}{M_{S}^2 e^{2w}} 
\right)\right] p(M_S) \nonumber
\end{eqnarray}
and $p(M_S)$ is a prior for $M_S$ itself, which we take to be
$p(M_S)=1/M_S$, i.e.\ flat in the logarithm of $M_S$. The marginalisation over
$M_S$ amounts to a marginalisation over a family of prior distributions, and
as such constitutes a hierarchical Bayesian approach~\cite{hierarchy}. The
integration over several distributions is equivalent to adding smearing due to
our uncertainty in the form of the prior. 
As far as we are aware, the present paper is the first example of the use of
hierarchical Bayesian techniques in particle physics. In general, we could
also have marginalised over the hyper-parameter $w$, for example using a
Gaussian centred on 1, but we find it useful below to examine sensitivity of
the posterior probability distribution to $w$. We therefore leave it as an
input parameter for the prior distribution.
We evaluate the integral in Eq.~\ref{priorsummary}
numerically using an integrator that does not evaluate the integrand at the
endpoints, where it is not finite. We have checked that the integral is not
sensitive to the endpoints chosen: the change induced by changing the
integration range
to $[10$ GeV, $10^{16}]$ GeV is negligible. 
We refer to
Eq.~\ref{priorsummary} as the ``same order'' prior.
To summarise, the posterior probability density function is given by
\begin{eqnarray}
p(m_0, M_{1/2}, A_0,\tan \beta, s | \mbox{data})
&\propto& \left[ p(\mbox{data} | m_0, M_{1/2}, A_0,\mu,B, s)
 \times  \label{summary} \right. \\ 
&& \left.
r(B, \mu, \tan \beta) \
p(s)\ p(m_0,M_{1/2},A_0, \mu,B)\right]_{M_Z=M_Z^{cen}},
 \nonumber 
\end{eqnarray}
where we have written
$\left[\dots\right]_{M_Z=M_Z^{cen}}$ on the right hand side of above relation,
  implying that $\mu$ and $B$ are eliminated in favour of $\tan \beta$ and
  $M_Z^{cen}$ by Eqs.~\ref{muB},~\ref{musq}.

\FIGURE[r]{\unitlength=1.1in
\begin{picture}(2.2,2.2)(0,0)
\put(0,0){\epsfig{file=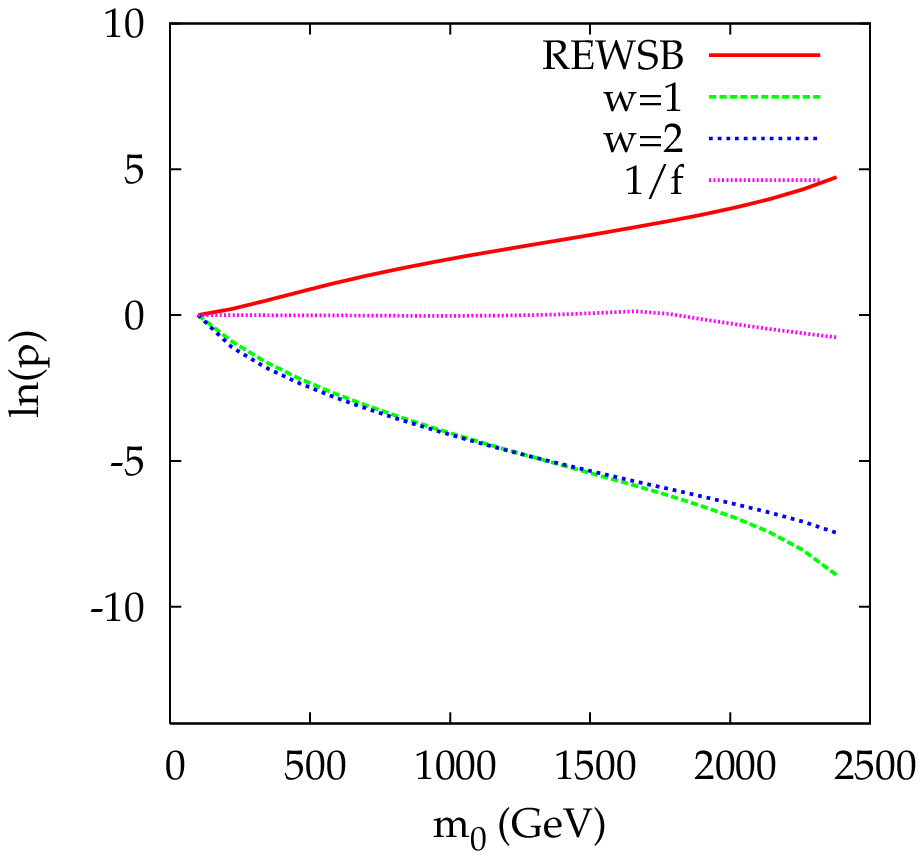, width=3in}}
\end{picture}
\caption{Prior factors $p$ in the CMSSM at SPS1a
  with varying $m_0$. Standard Model inputs have been fixed at their
  empirically central 
values. \label{fig:ft}}} 
We may view the prior factors in Eq.~\ref{summary} to be inverse fine-tuning
parameters: where the fine-tuning is high, the priors are
small. It is
interesting to note that a cancellation of order $\sim 1/ \tan \beta$ is
known to be required in order to achieve high values of $\tan
\beta$~\cite{hightanb}. This appears in our Bayesian prior as a result of
transforming from the fundamental Higgs potential parameters $\mu$, $B$ to
$\tan \beta$ and the empirically preferred value of $M_Z$. 
We display the various prior factors in Fig.~\ref{fig:ft} as a function of
  $m_0$ for all other parameters at the SPS1a CMSSM
  point~\cite{Allanach:2002nj}: $M_{1/2}=250$ GeV, $A_0=100$ GeV, $\tan
  \beta=10$ and all SM input parameters fixed at their central empirical
  values. The figure displays the REWSB prior, the REWSB prior+same order
  priors with $w=1, 2$ (simply marked $w=1$, $w=2$ respectively) and
  the inverse of the fine-tuning parameter defined in Eq.~\ref{ft}.
  We see that the REWSB prior actually increases with $m_0$ along the
  chosen line in CMSSM parameter space. This is due to decreasing $\mu$ in
  Eq.~\ref{rewsb} 
  towards the focus-point\footnote{The focus-point region is a subset of the
  hyperbolic branch~\protect\cite{hyper}.} at high
  $m_0$~\cite{Feng:1999zg}. The   conventional 
  fine-tuning measure $f$ remains roughly constant as a function of $m_0$,
  whereas the same order priors decrease strongly as a function of $m_0$. This
  is driven largely by the $1/m_0$ factor in Eq.~\ref{priorsummary} and
  the mismatch between large $m_0$ and $M_{1/2}=250$
  GeV, which leads to a  stronger suppression for the smaller width $w=1$
  rather than $w=2$.

The SM input parameters $s$ used are displayed in Table~\ref{tab:sminp}.
Since they have all been well measured, their priors are set to be Gaussians
with central values and widths as listed in the table. 
We use Ref.~\cite{pdg} for the QED coupling
constant $\alpha^{\overline{MS}}$, the strong coupling constant
$\alpha_s^{\overline{MS}}(M_Z)$ and the running mass of the bottom quark 
$m_b(m_b)^{\overline{MS}}$, all in the $\overline{MS}$ renormalisation scheme.
A recent Tevatron top mass $m_t$ measurement \cite{mtmeas} is also
employed, although the absolutely latest value has shifted slightly~\cite{latemt}. 
$p(s)$ is set to be a product of Gaussian probability
distributions\footnote{Taking the product corresponds to assuming that the
  measurements are independent.} 
$p(s) \propto \prod_i e^{-\chi^2_i}$, where 
\begin{equation}
\chi_i^2 = \frac{(c_i - p_i)^2}{\sigma_i^2} \label{chisq}
\end{equation}
for observable $i$.  $c_i$ denotes the central value of the
experimental measurement, $p_i$ represents the value 
of SM input parameter $i$.
Finally $\sigma_i$ is the standard error of the measurement.  

\TABULAR[r]{|c|c|}{\hline
SM parameter & constraint \\ \hline
$1/\alpha^{\overline{MS}}$ & 127.918$\pm$0.018 \\
$\alpha_s^{\overline{MS}}(M_Z)$ & 0.1176$\pm$0.002 \\
$m_b(m_b)^{\overline   MS}$ & 4.24$\pm$0.11 GeV \\
$m_t$ & 171.4$\pm$2.1 GeV \\ \hline
}{SM input parameters \label{tab:sminp}}
We display marginalised prior pdfs in
Fig.~\ref{fig:priors} for the REWSB, REWSB+same order ($w=1$) and
REWSB+same order ($w=2$) priors. The plots have 75 bins and 
the prior pdf has been marginalised over all unseen
dimensions. No indirect data has been taken into account in producing the
distributions, a feasible electroweak symmetry breaking vacuum being the only
constraint. 
The priors have been obtained by sampling with a MCMC using the
Metropolis 
algorithm~\cite{metropolis,mackay}, taking the average of 10 chains of 100~000
steps each.
Figs.~\ref{fig:priors}a,b shows that although the same order priors are
heavily peaked 
towards small values of $m_0<500$ GeV and $M_{1/2} \sim 180$ GeV,
the 95$\%$ upper limits shown by the vertical arrows are only moderately
constrained for $m_0$. $w=1$ is not surprisingly more peaked at lower mass
values. The REWSB histograms on the other hand, prefer high $m_0$ (due to the
lower values of $\mu$ there) and are quite flat in $M_{1/2}$. The same order
of magnitude requirement is crucial in reducing the preferred scalar masses.
The REWSB prior is fairly flat in $A_0$ whereas the $w=1$, 
$w=2$ priors are heavily peaked around zero.
The $M_{1/2}$ same-order priors are more strongly peaked than, for example,
$m_0$ because $M_{1/2}$ is strongly correlated with $|\mu|$ and so the
logarithmic measure of the prior (leading to the factor of $1/(m_0 M_{1/2}
|\mu|)$ in Eq.~\ref{priorsummary} becomes more strongly suppressed.
 $\tan \beta$ is
peaked very strongly toward lower values of the considered range
for the REWSB prior due to the $1/\tan \beta$ suppression, but becomes
somewhat diluted when the same order priors 
are added, as shown in Fig.~\ref{fig:priors}d. 
\FIGURE{\fourgraphs{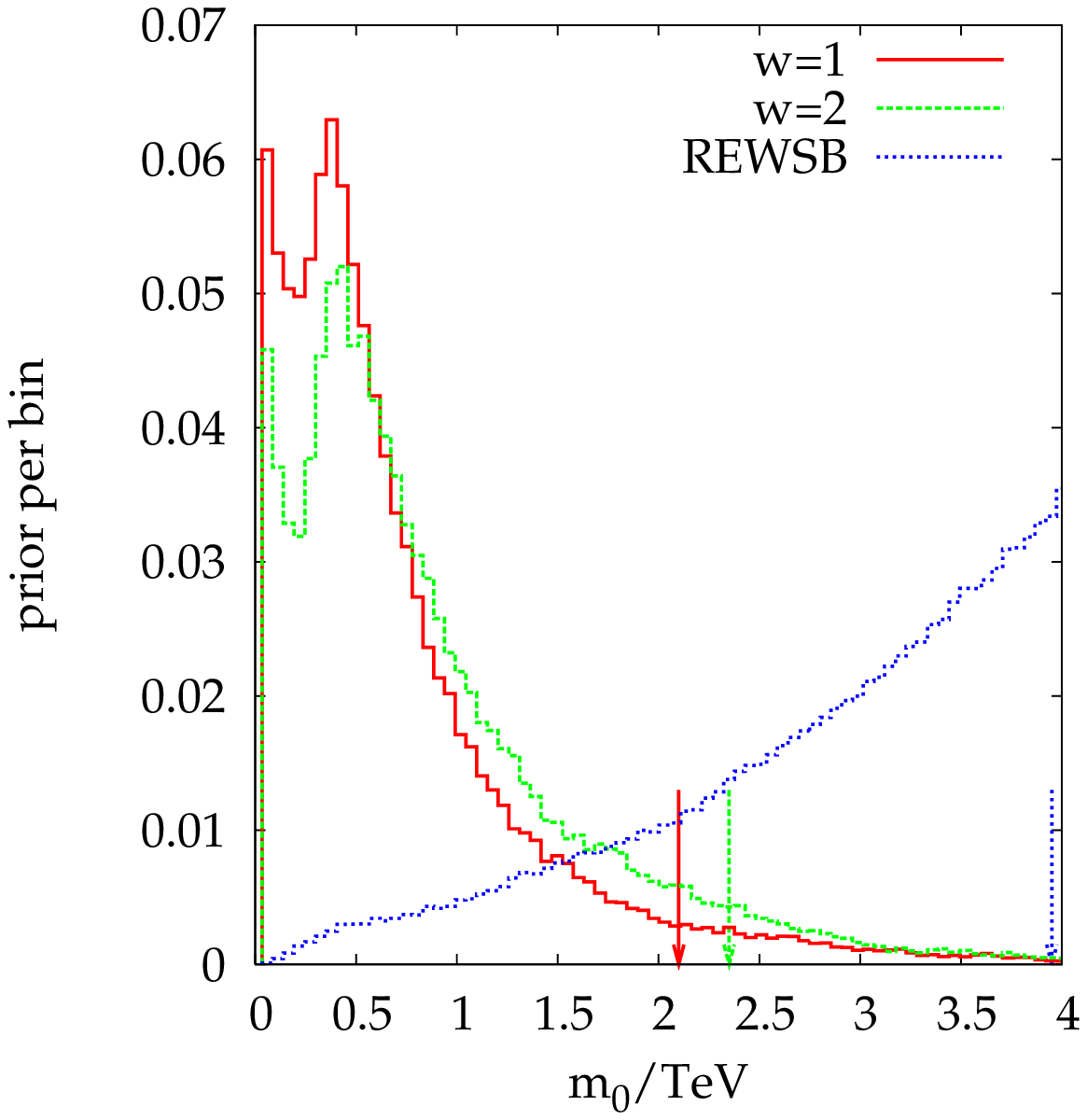}{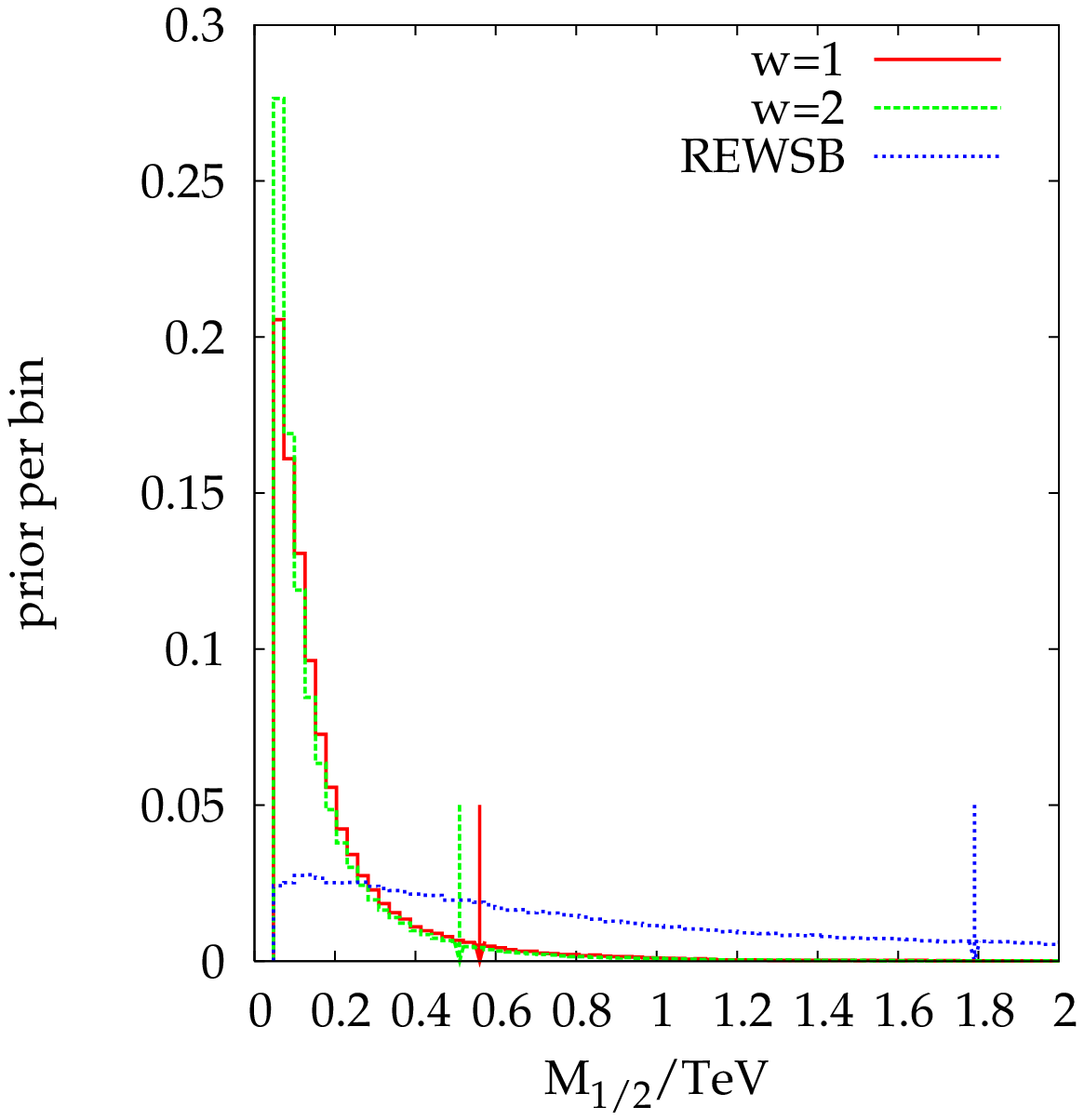}{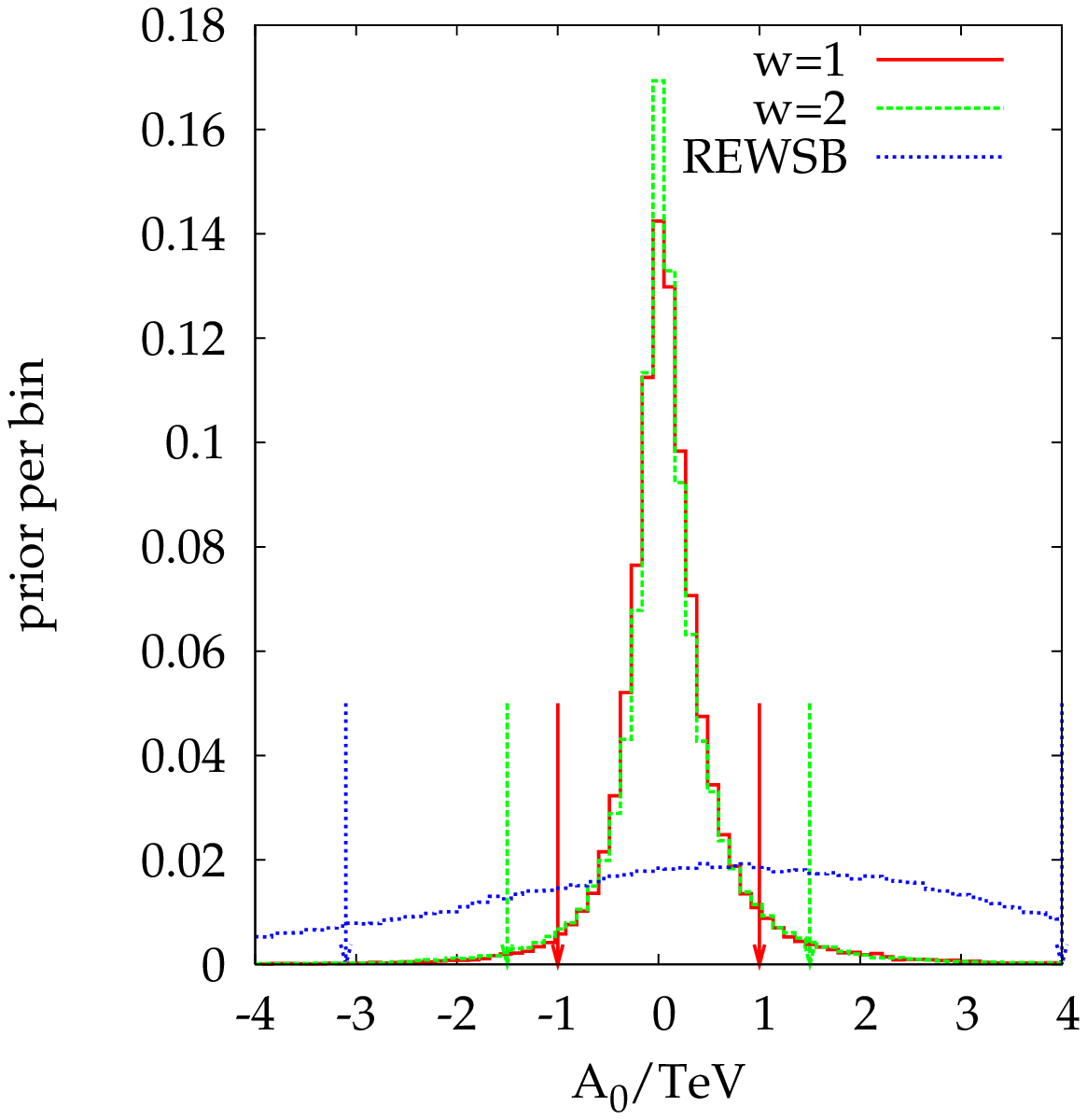}{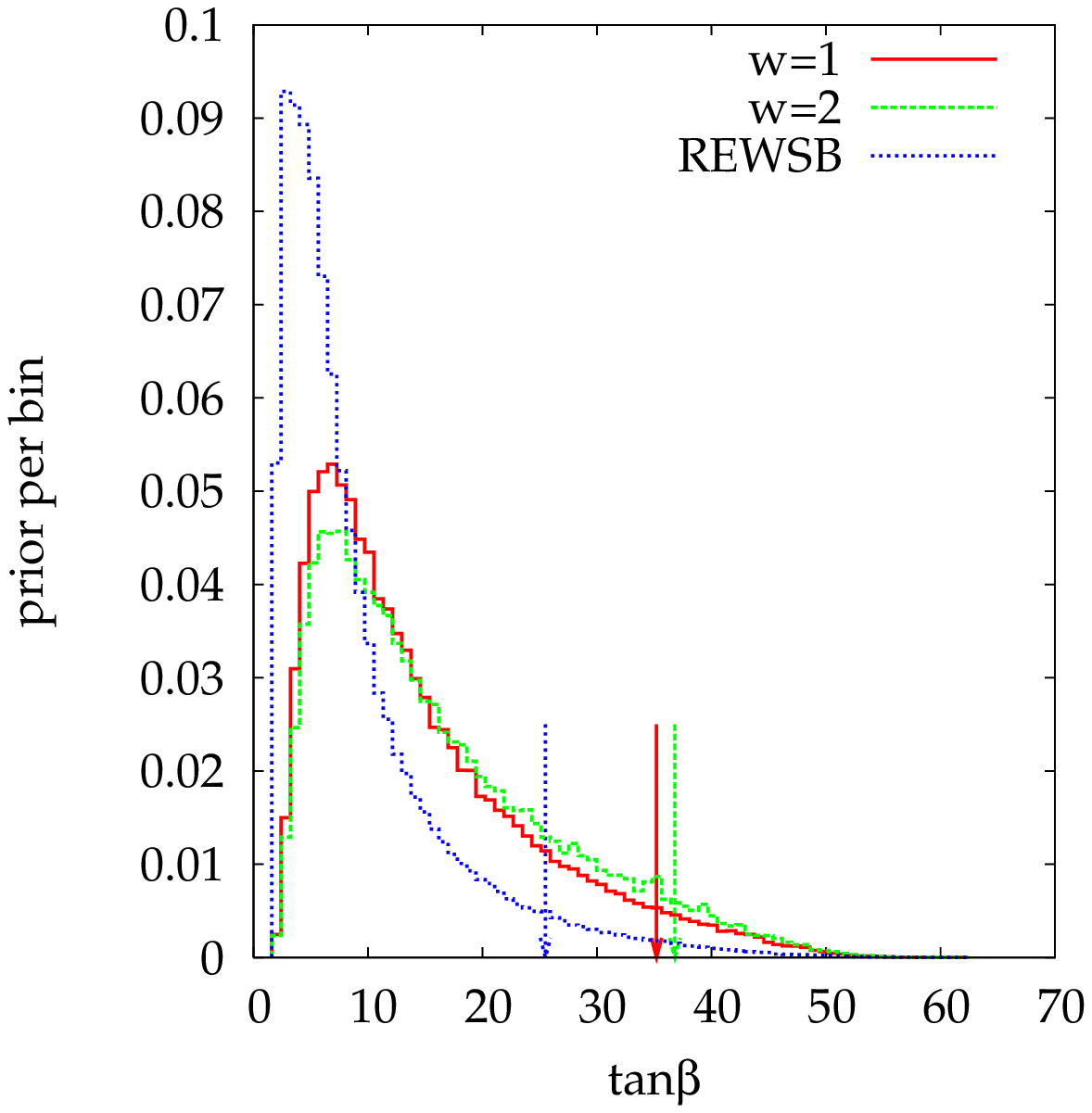}
\caption{Prior probability distributions marginalised to 
the (a) $m_0$,
 (b) $M_{1/2}$, (c) $A_0$ and (d) $\tan \beta$ directions. 95$\%$ upper limits
are shown by the labelled arrows except in (c), where the arrows delimit the
2-sided 95$\%$ confidence region. All distributions have been binned with 75
equally spaced bins.
\label{fig:priors}}}

\section{The Likelihood \label{sec:likelihood}}
\TABULAR[r]{|c|c|}{\hline
CMSSM parameter & range \\ \hline
$A_0$ & -4 TeV to 4 TeV \\
$m_0$ & 60 GeV to 4 TeV \\
$M_{1/2}$ & 60 GeV to 2 TeV \\
$\tan \beta$ &  2 to 62 \\ \hline
}{Input parameters \label{tab:inp}}
Our calculation of the likelihood closely follows
Ref.~\cite{Allanach:2006jc}. For completeness, we describe the procedure
here. 
Including the SM inputs in Table~\ref{tab:sminp}, eight input parameters
are varied simultaneously. 
The range of CMSSM parameters considered is shown in
Table~\ref{tab:inp}. The SM 
input parameters are allowed to
vary within 4$\sigma$ of their central values. Experimental errors are so
small on  the muon decay constant $G_\mu$ that we 
fix it to its central value of $1.16637 \times 10^{-5}$
GeV$^{-2}$.

In order to calculate predictions for observables from the inputs, the program 
{\tt SOFTSUSY2.0.10}~\cite{softsusy} is 
first employed to calculate the MSSM spectrum.
Bounds upon the sparticle spectrum have been updated and are based upon the bounds
collected in Ref.~\cite{deAustri:2006pe}.
Any spectrum violating a 95$\%$ limit from negative
sparticle searches is assigned a zero likelihood density. 
Also, we set a zero likelihood for any inconsistent point, e.g.\ one which does
not break electroweak symmetry correctly, or a point that contains tachyonic
sparticles. 
For points that are not ruled out,
we then link the MSSM spectrum via the SUSY Les Houches Accord~\cite{slha} to
  {\tt micrOMEGAs1.3.6}~\cite{micromegas}, which then calculates $\Omega_{DM}
  h^2$, the
branching ratios $BR(b \rightarrow s \gamma)$ and $BR(B_s \rightarrow \mu^+
\mu^-)$ and the anomalous magnetic moment of the muon $(g-2)_\mu$.  

The anomalous magnetic moment of the muon $a_\mu\equiv(g-2)_\mu/2$ 
was measured to be $a^\mathrm{exp}_\mu=(11659208.0\pm5.8)\times 10^{-10}$ \cite{gm2exp}.
Its experimental value is in conflict with the SM predicted value  
$a_\mu^{\mathrm{SM}}=(11659180.4\pm5.1)\times10^{-10}$  from~\cite{gm2SM}, 
which comprises the latest QED~\cite{gm2QED}, electroweak~\cite{gm2EW},
 and hadronic~\cite{gm2SM} contributions to $a^{\mathrm{SM}}_\mu$.
This SM prediction however does not account for $\tau$ data which is known 
to lead to significantly different results for $a_\mu$, 
implying underlying theoretical difficulties which have not been resolved so far.
Restricting to $e^+e^-$ data, hence using the numbers given above, we find
\begin{equation}
  \delta \frac{(g-2)_\mu}{2}\equiv \delta a_\mu 
 \equiv a_\mu^{\mathrm{exp}}-a_\mu^{\mathrm{SM}}=(27.6 \pm 7.7)\times 10^{-10}.
\end{equation}
This excess may be explained by a supersymmetric contribution,
the sign of which is identical to the sign of the superpotential $\mu$
parameter~\cite{susycont}. After obtaining the one-loop MSSM value of
$(g-2)_\mu$ from {\tt micrOMEGAs1.3.6}, we add the dominant 2-loop
corrections detailed in Refs.~\cite{2loop,private}.
The $W$ boson mass $M_W$ and the 
effective leptonic mixing angle $\sin^2 \theta^l_w$
are also used in the likelihood. We take the 
measurements to be~\cite{mw,sinth} 
\begin{equation}
M_W = 80.398\pm0.027\mbox{~GeV},
\qquad \sin^2 \theta_w^l = 0.23153 \pm 0.000175, \label{ewobs}
\end{equation}
where experimental errors and theoretical uncertainties due to missing 
higher order corrections in SM~\cite{mwsmbest} 
and MSSM~\cite{Heinemeyer:2006px,drMSSMal2B} have been added in quadrature. 
The most up to date MSSM predictions for 
$M_W$ and $\sin^2 \theta_w^l$~\cite{Heinemeyer:2006px}  
are finally used to compute the corresponding likelihoods.
A parameterisation of the LEP2 Higgs search likelihood for
various Standard Model Higgs masses is utilised, since the lightest Higgs $h$
of the CMSSM is very SM-like once the direct search constraints are taken into
account. It is smeared with a 2 GeV assumed theoretical uncertainty in the {\tt
  SOFTSUSY2.0.10} prediction of $m_h$ as described in
Ref.~\cite{Allanach:2006jc}. 
The rare bottom quark branching ratio to a strange quark and a photon 
$BR(b \rightarrow s \gamma)$ is constrained to be~\cite{hfg}
\begin{equation}
BR(b \rightarrow s \gamma)=  (3.55\pm0.38) \times 10^{-4}, \label{bsg}
\end{equation}
obtained by adding the experimental error with the estimated theory
error~\cite{gamb} of $0.3 \times 10^{-4}$ in quadrature.
The WMAP3~\cite{wmap} power law $\Lambda$-cold dark matter fitted value of the dark matter
relic density is
\begin{equation}
\Omega \equiv \Omega_{DM} h^2 = 0.104^{+0.0073}_{-0.0128} \label{omega}
\end{equation}
In the present paper, we assume that all of the dark matter consists of
neutralino lightest supersymmetric particles and we enlarge the errors on
$\Omega_{DM} h^2$ to $\pm 0.02$ 
in order to incorporate an estimate of higher order uncertainties in its
prediction. 

We assume that the measurements and thus also the likelihoods extracted from
$\Omega$,  
$BR(b \rightarrow s \gamma)$, $M_W$, $\sin^2 \theta_w^l$, $(g-2)_\mu$, $BR(B_s
\rightarrow \mu^+
\mu^-)$ are all independent of each other so that the individual likelihood
contributions may be multiplied. 
Observables that have been quoted with
uncertainties are assumed to be Gaussian distributed and are 
characterised by $\chi^2$. 

\section{CMSSM Fits With the New Priors \label{sec:fits}}

In order to sample the posterior probability density, we ran 10 independent
MCMCs of 500~000
steps each using a newly developed banked~\cite{bank} 
Metropolis-Hastings MCMC\@. The banked method was specifically designed to 
sample
several well isolated or disconnected local maxima, for example maxima in the
posterior pdfs of $\mu>0$ and $\mu<0$. Previously, we had normalised the two
samples via bridge sampling~\cite{darkSide}, which requires  twice the
number of samples than for one maximum, with additional calculations
required after the sampling. Bank sampling, on the other hand, can be performed
with roughly an identical number of sampling steps to the case of one maximum
and does not require additional normalisation calculations after the sampling.
The chance of a bank proposal for the position of
the next point in the chain was set to 0.1, meaning that the usual Metropolis
proposal had a chance of 0.9.
The bank was formed from 10 initial Metropolis MCMC
runs with 60~000 steps each and random starting points that were drawn from
pdfs flat in the ranges displayed in Tables~\ref{tab:sminp},\ref{tab:inp}. The
initial 4000 
steps were discarded in order to provide adequate ``burn-in'' for the MCMCs.
We check convergence using the Gelman-Rubin $\hat{R}$
statistic~\cite{gelman,Allanach:2005kz}, which provides an estimated upper
bound on how much the variance in 
parameters could be decreased by running for more steps in the chains. Thus,
values close to 1 show convergence of the chains. In previous publications, we
considered 
$\hat{R}<1.05$ to indicate convergence of the chains for every input
parameter. We have checked that this is easily satisfied for all of our
results. 

We compare the case of flat $\tan \beta$ priors to the new prior in
Fig.~\ref{fig:newprior}. The posterior pdf has been marginalised down to the
$M_{1/2}-m_0$ plane and binned into 75$\times$75 bins, as with all
two-dimensional distributions in the present paper. Both signs of $\mu$ have
been marginalised over, again like all following figures in this paper unless
explicitly mentioned. 
The bins are normalised
with respect to the bin with maximum posterior. 
\FIGURE{\fourgraphst{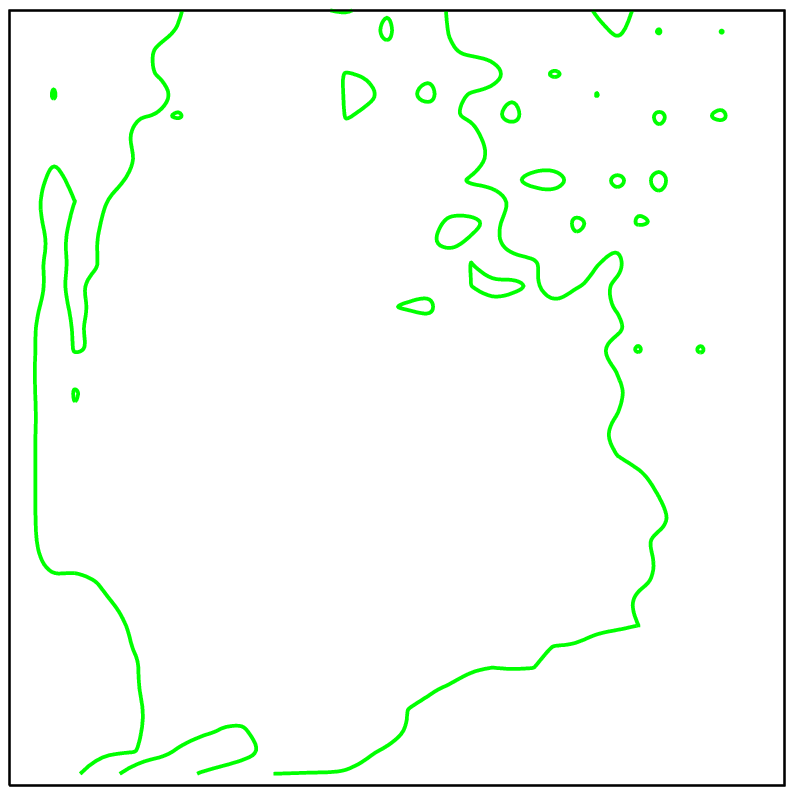}{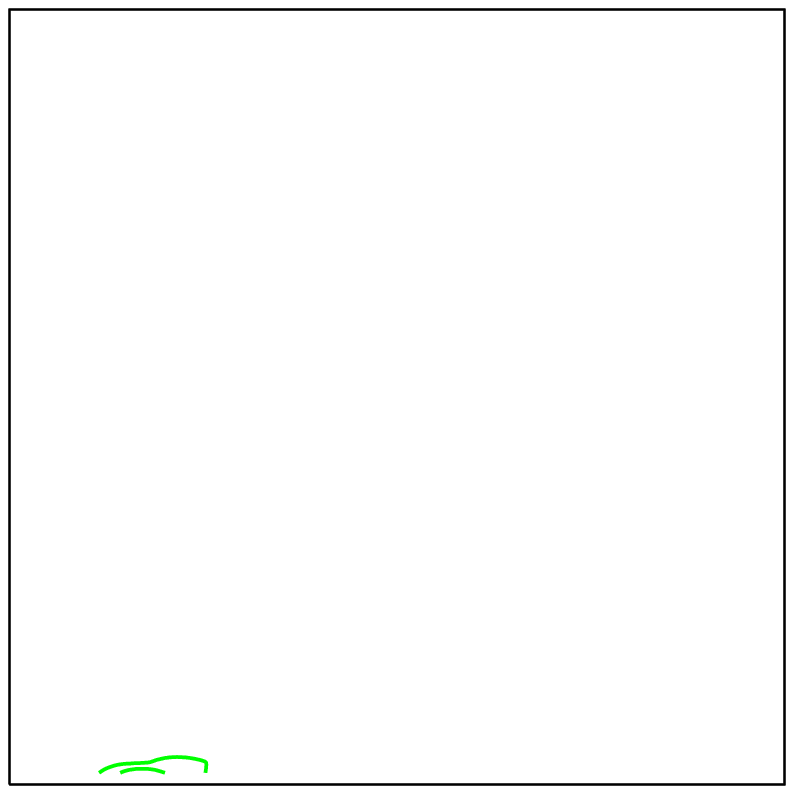}{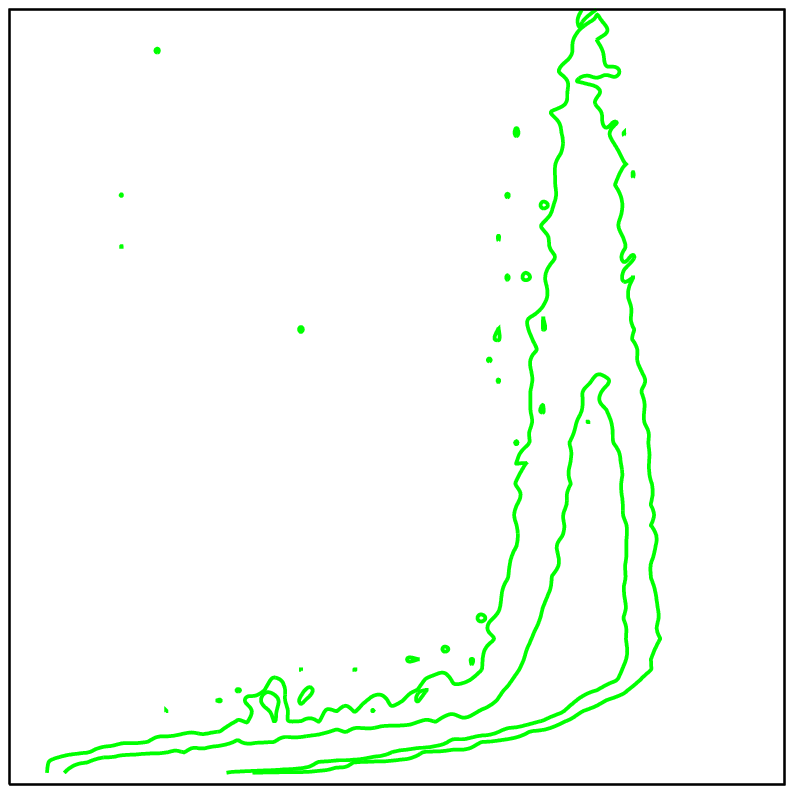}{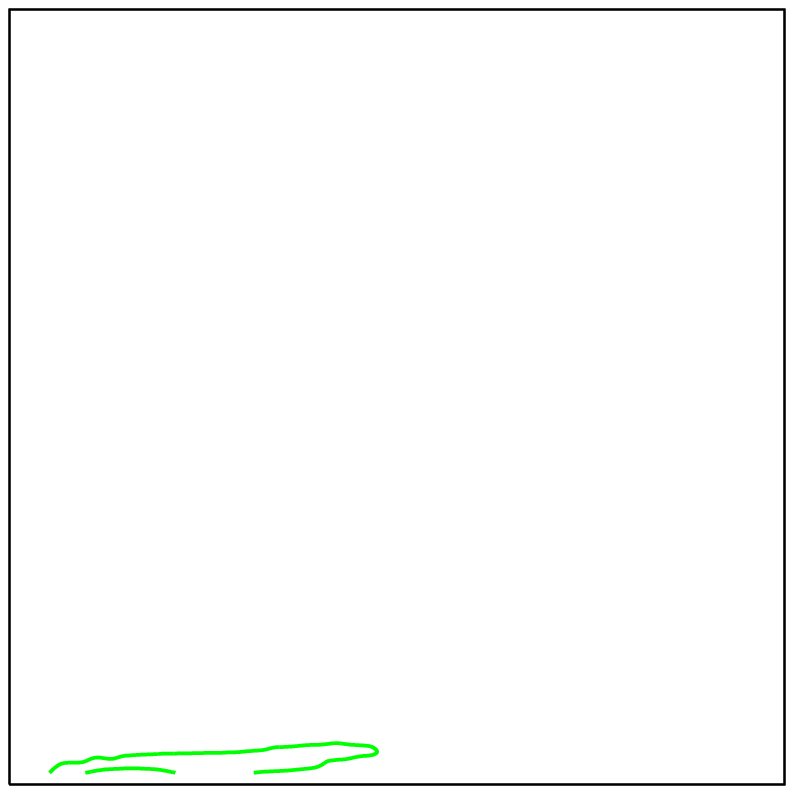}
\caption{CMSSM fits marginalised in the unseen dimensions for
(a,c) flat $\tan \beta$ priors, (b,d) the REWSB+same order prior
with $w=1$. Contours showing the
  68$\%$   and 95$\%$ regions are shown in each case. The posterior
  probability in each bin, normalised to the probability of the maximum bin,
  is displayed by reference to the colour bar on the 
  right hand side of each plot.
\label{fig:newprior}}}
We identify the usual CMSSM regions of good-fit in
Fig.~\ref{fig:newprior}a. The maximum 
at the lowest value of $m_0$ corresponds to the stau co-annihilation region~\cite{Griest:1990kh},
where ${\tilde \tau}_1$ and $\chi_1^0$ are quasi-mass degenerate and
efficiently annihilate in the early universe. This region is associated with 
$\tan \beta<40$, as Fig.~\ref{fig:newprior}b indicates. $m_0
\sim 1$ TeV in Fig.~\ref{fig:newprior}a has large $\tan \beta \sim 50$. 
This region corresponds to the case where the neutralinos efficiently
annihilate through $s-$channel pseudoscalar Higgs bosons $A^0$ into $b
\bar{b}$ and $\tau \bar{\tau}$ pairs~\cite{Drees:1992am,Arnowitt:1993mg}. The
region at low $M_{1/2}$ and 
high $m_0$ in Fig.~\ref{fig:newprior}a is the $h^0$ pole
region~\cite{Djouadi:2005dz},
where neutralinos annihilate predominantly through an $s-$channel of the
lightest  CP even  Higgs $h^0$. In order to evade LEP2 Higgs constraints, this
also requires large $\tan \beta$. The focus point
region~\cite{Feng:1999mn,Feng:1999zg,Feng:2000gh} is the region around 
$M_{1/2}\sim 0.5$ TeV and $m_0=2-4$ TeV, where the lightest neutralino has a
significant higgsino component, leading to efficient annihilation into gauge
boson pairs. This region is somewhat sub-dominant in the fit, but extends
through most of the range of $\tan \beta$ considered. 

We see a marked difference between Figs.~\ref{fig:newprior}a
and~\ref{fig:newprior}b.  
The $A^0$ and $h^0$ pole regions have vanished with the
REWSB priors. The $A^0$ pole region is suppressed because the REWSB
prior disfavours the required large values of 
$\tan \beta$, as shown in Fig.~\ref{fig:priors}d. The $h^0$ pole region is
suppressed because the REWSB prior disfavours large values of $|A_0|$,
see 
Fig.~\ref{fig:priors}c,  and
large values of $|A_0|/M_{1/2}$. Large values of $|A_0|$ are necessary in this
region 
in order to achieve large stop mass splitting and therefore large corrections
to the lightest Higgs mass. Without such corrections, $h^0$ falls foul of
LEP2 Higgs mass bounds. 
The focus-point region has
been diminished by the REWSB priors mainly because the large values of
$m_0$ required become suppressed as in Fig.~\ref{fig:priors}a. This
suppression comes primarily from the requirement that SUSY breaking and Higgs
parameters be roughly of the same order as each other.
Figs.~\ref{fig:newprior}b,d display only one good-fit region corresponding to
the stau co-annihilation region at low $m_0$. 
The banked method~\cite{bank} allows an efficient normalisation of
the $\mu>0$ and $\mu<0$ branches, both of which are included in the figure. 

We now turn to a comparison of the REWSB+same order prior fits. We
consider such fits to give much more reliable results than the flat $\tan
\beta$ fits, and a large difference between fits for $w=1$ to $w=2$
would provide evidence for a lot of sensitivity to our exact choice of prior. 
Some readers might consider the flat $\tan \beta$ priors to be not
unreasonable, and those readers could take the large difference between flat
priors and the new more natural ones as a result of uncertainty 
originating from scarce data. 
\FIGURE{\eightgraphs{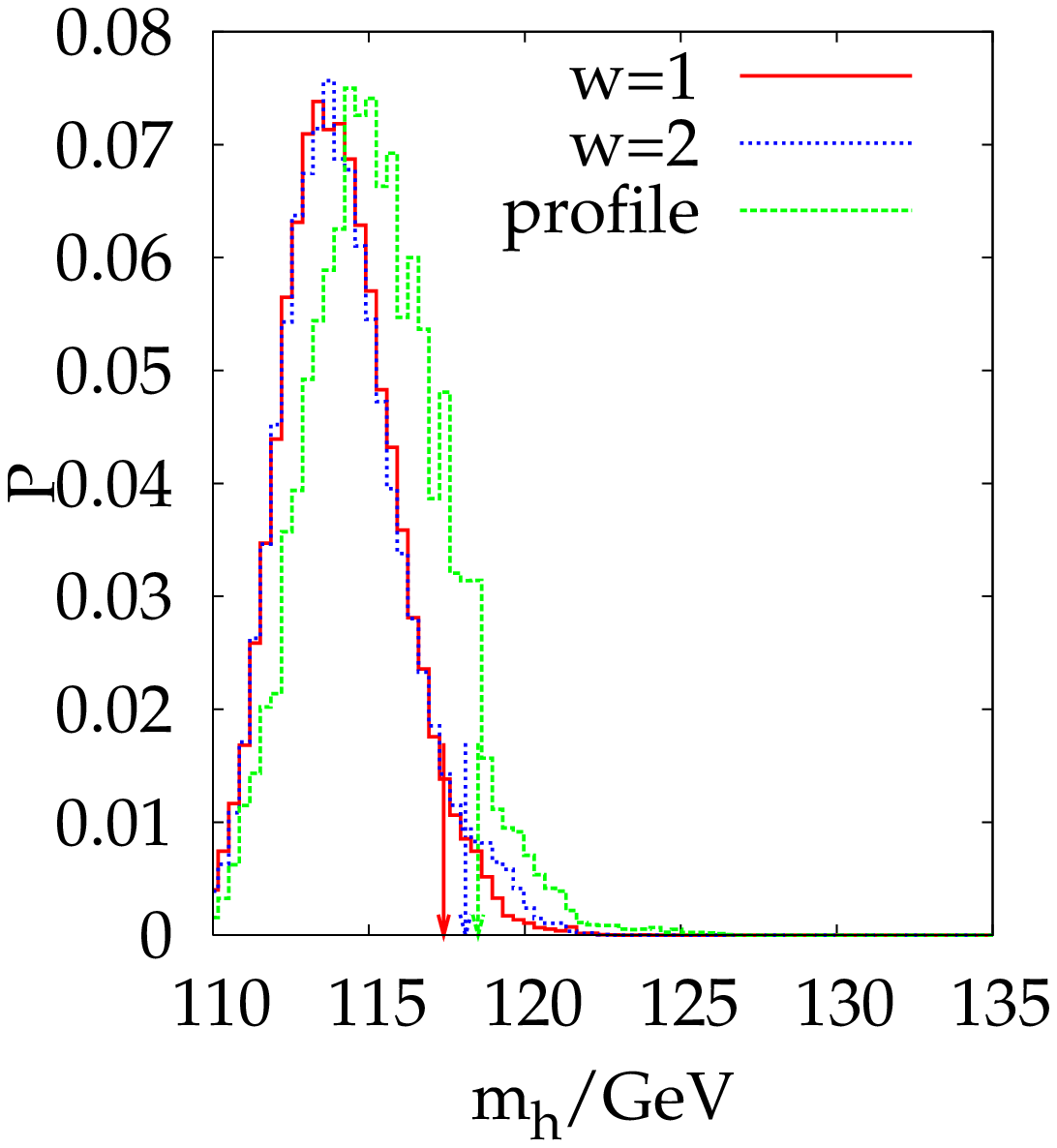}{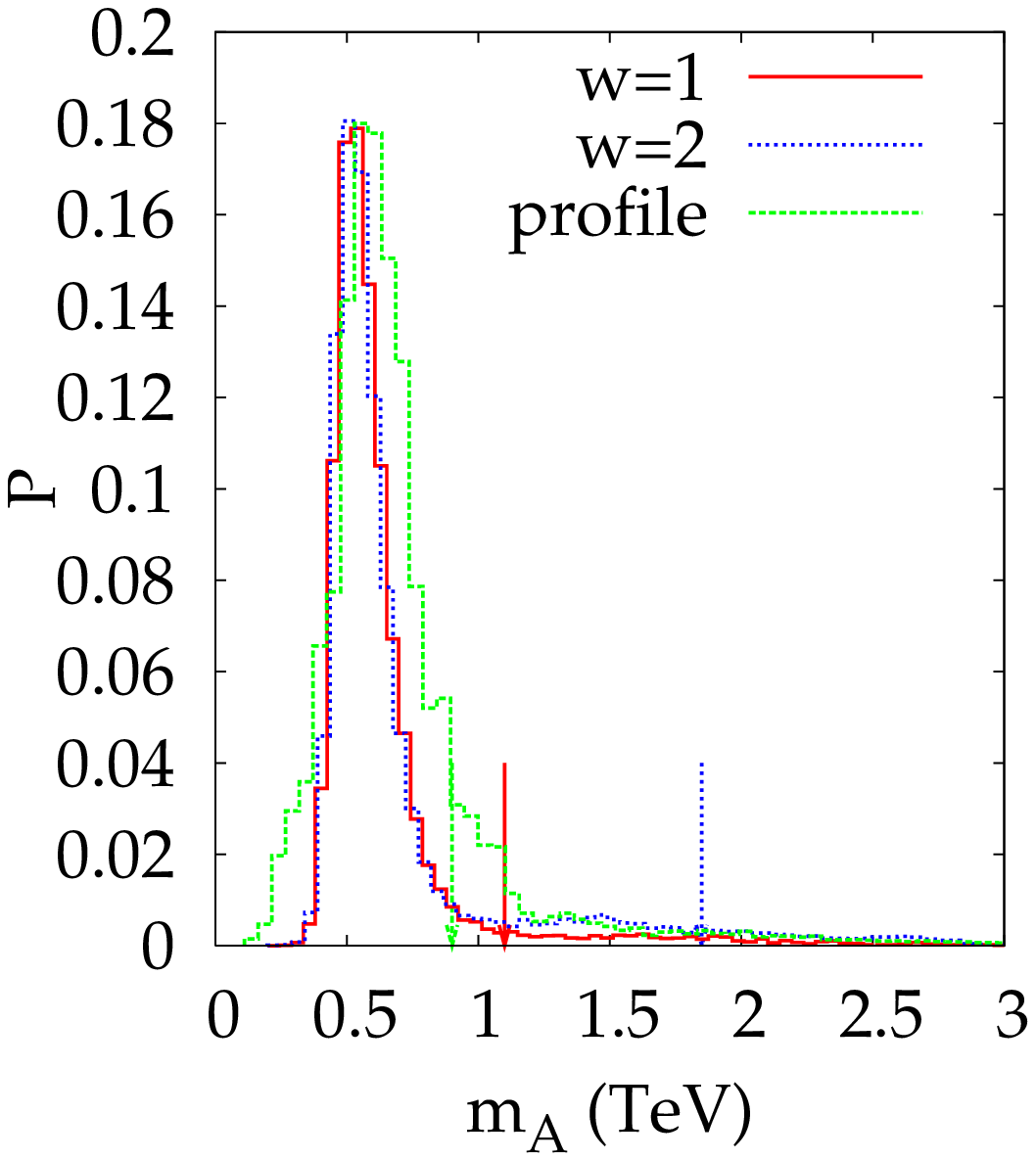}{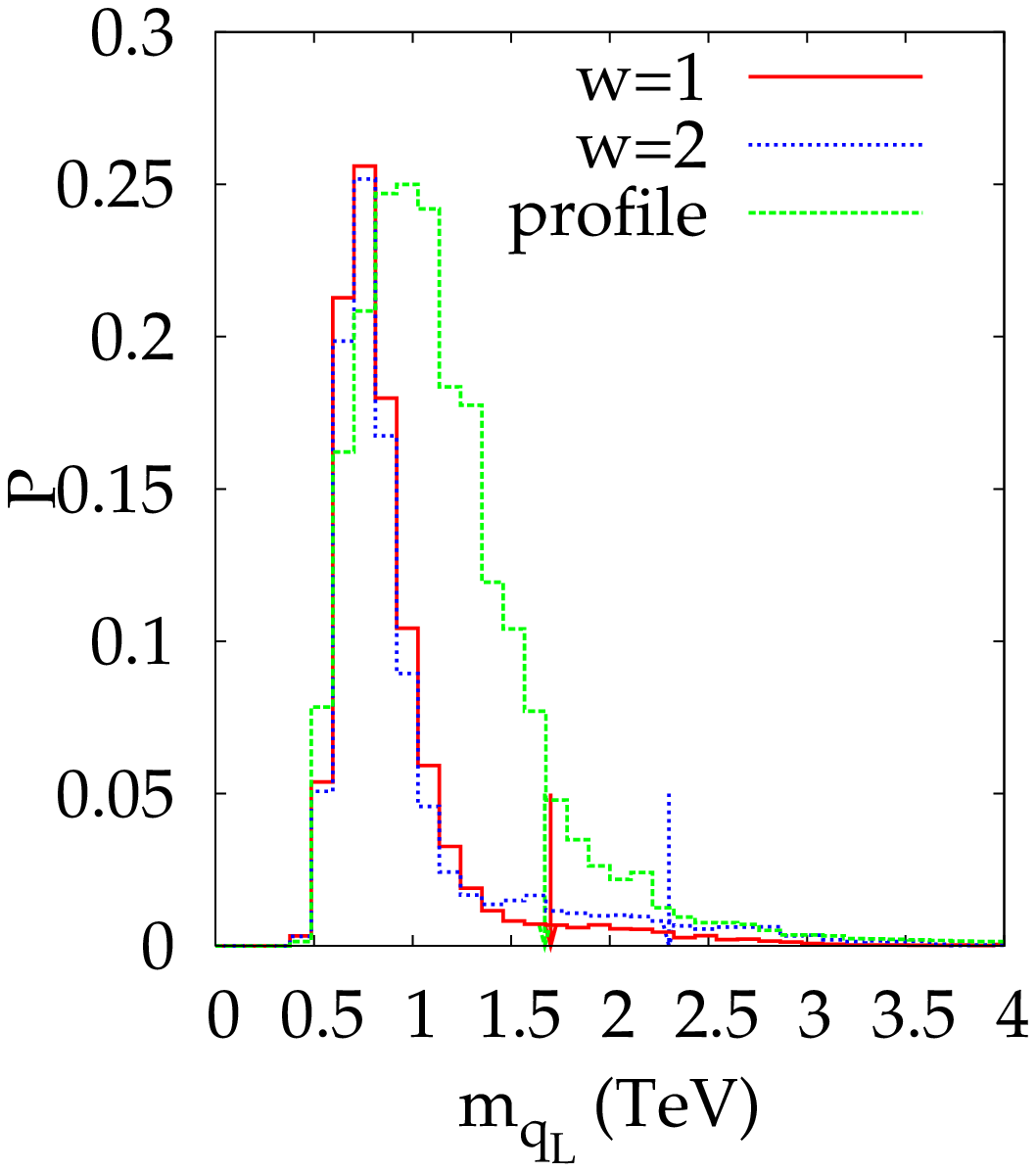}{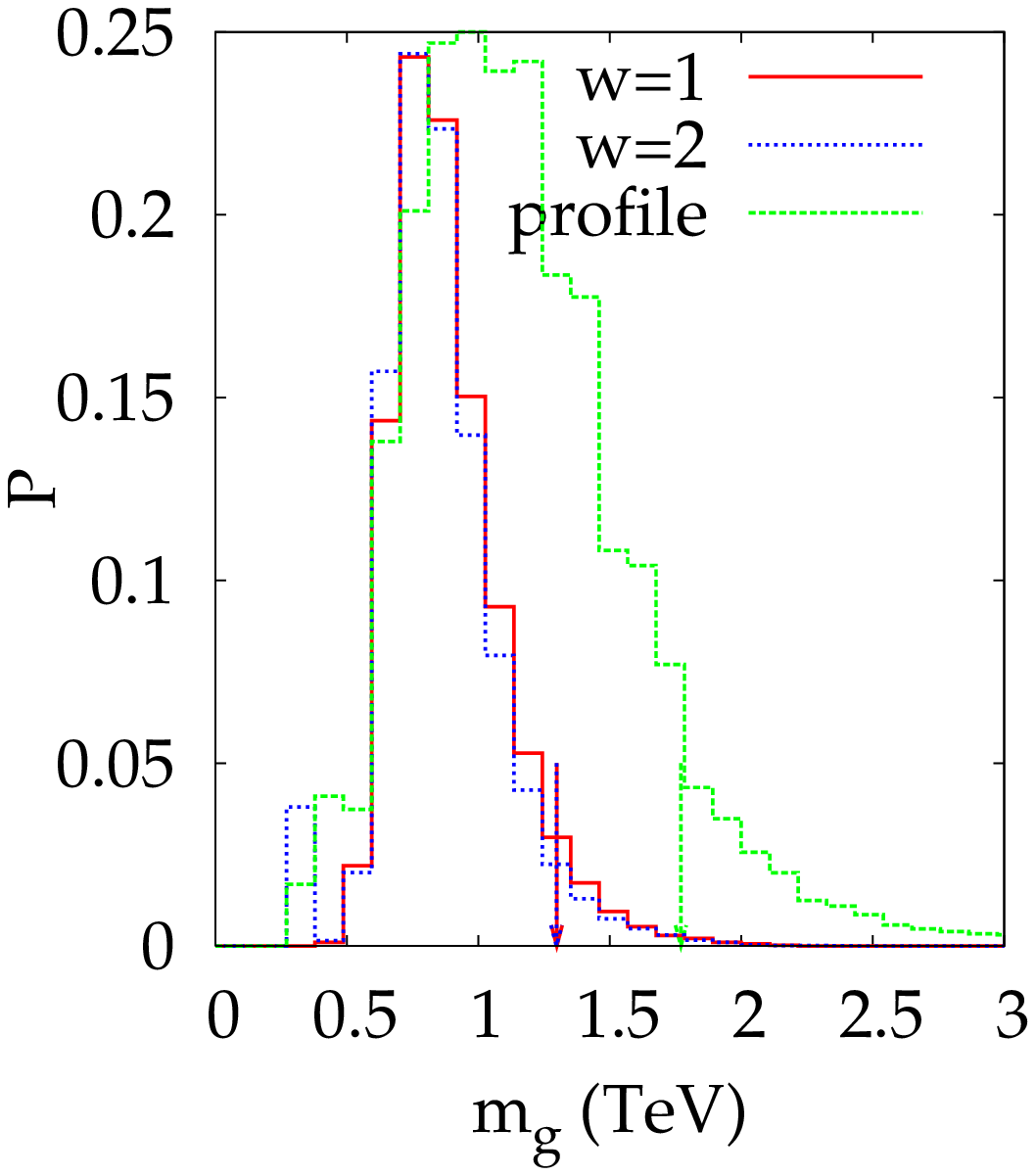}{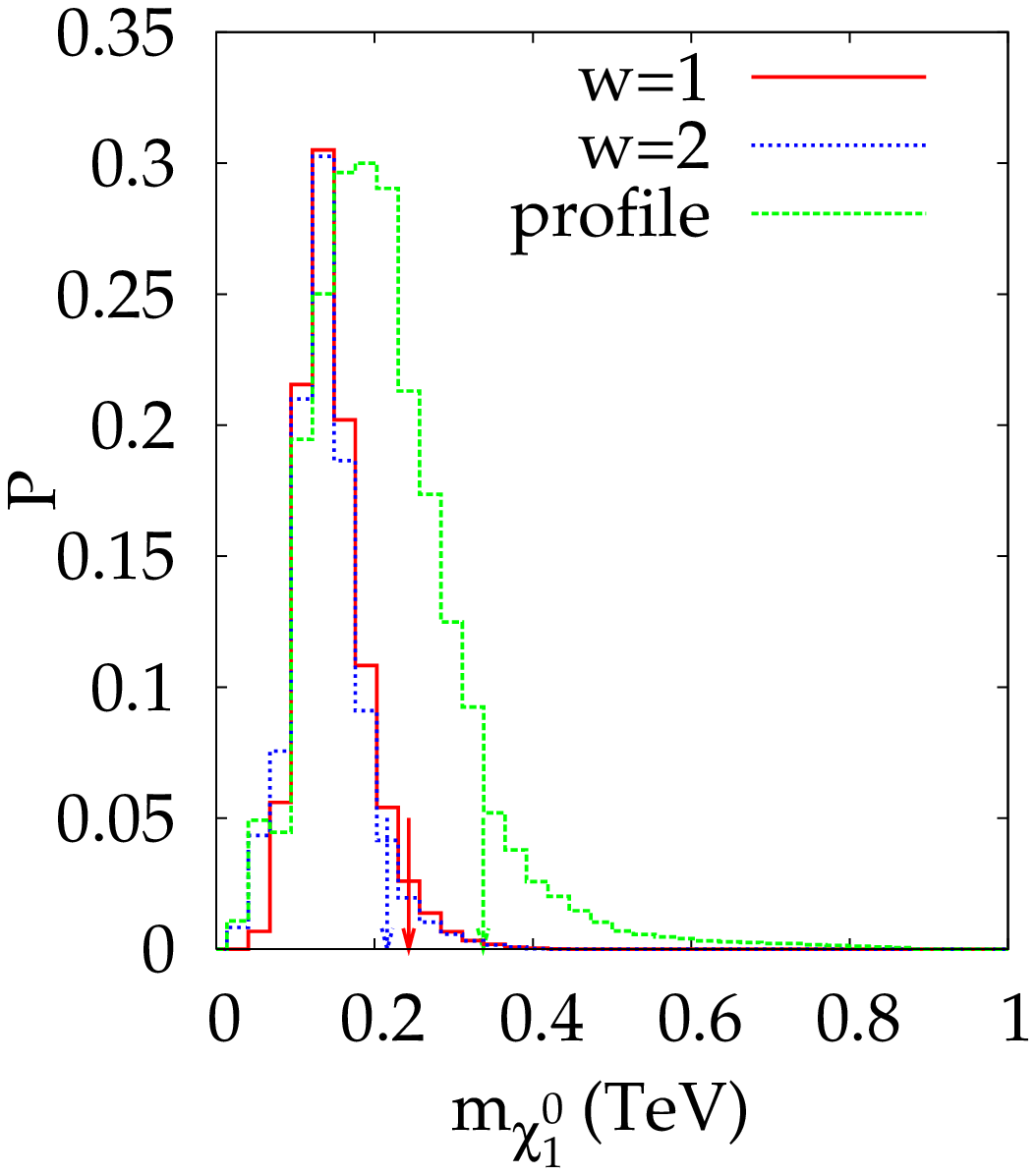}{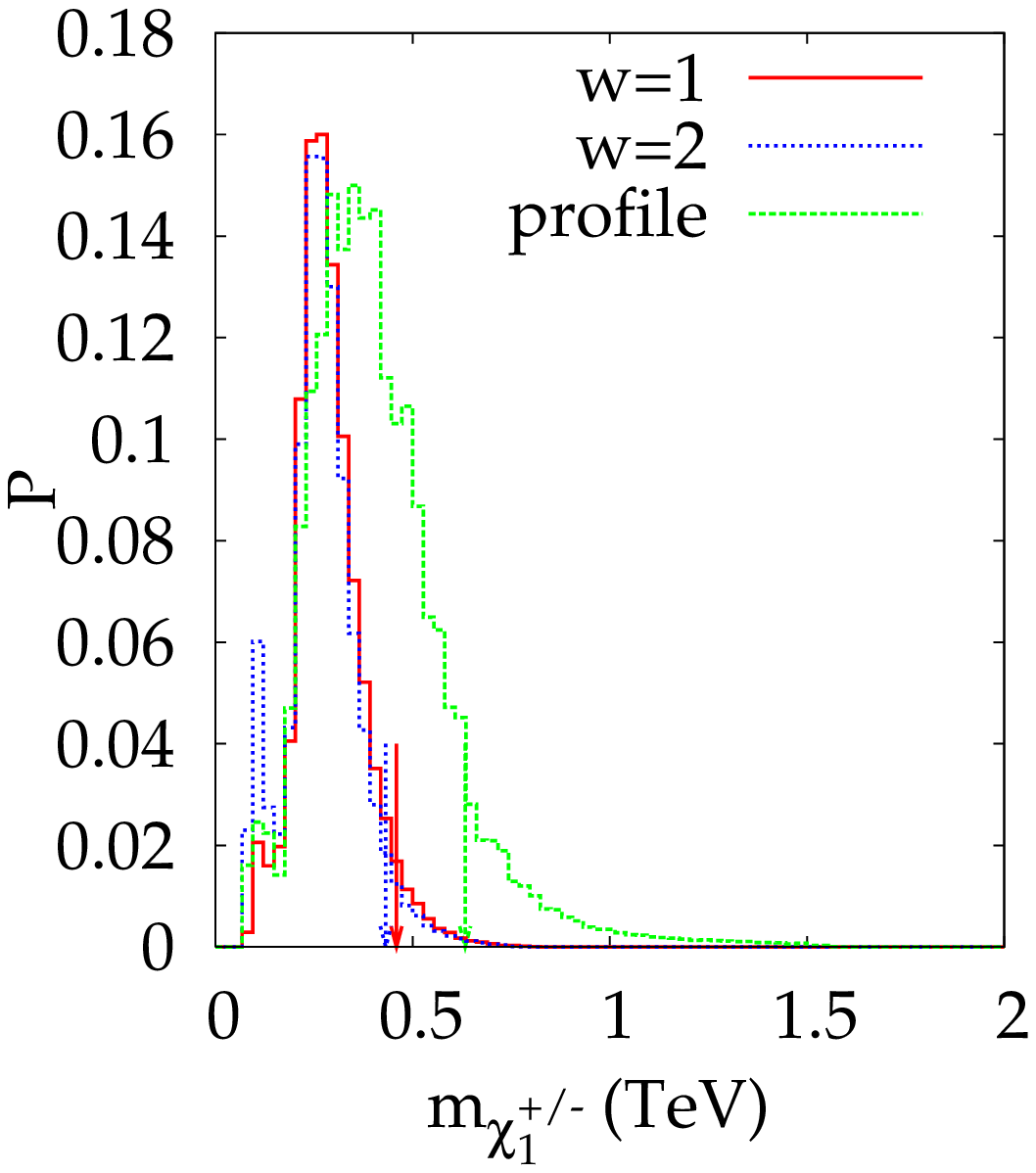}{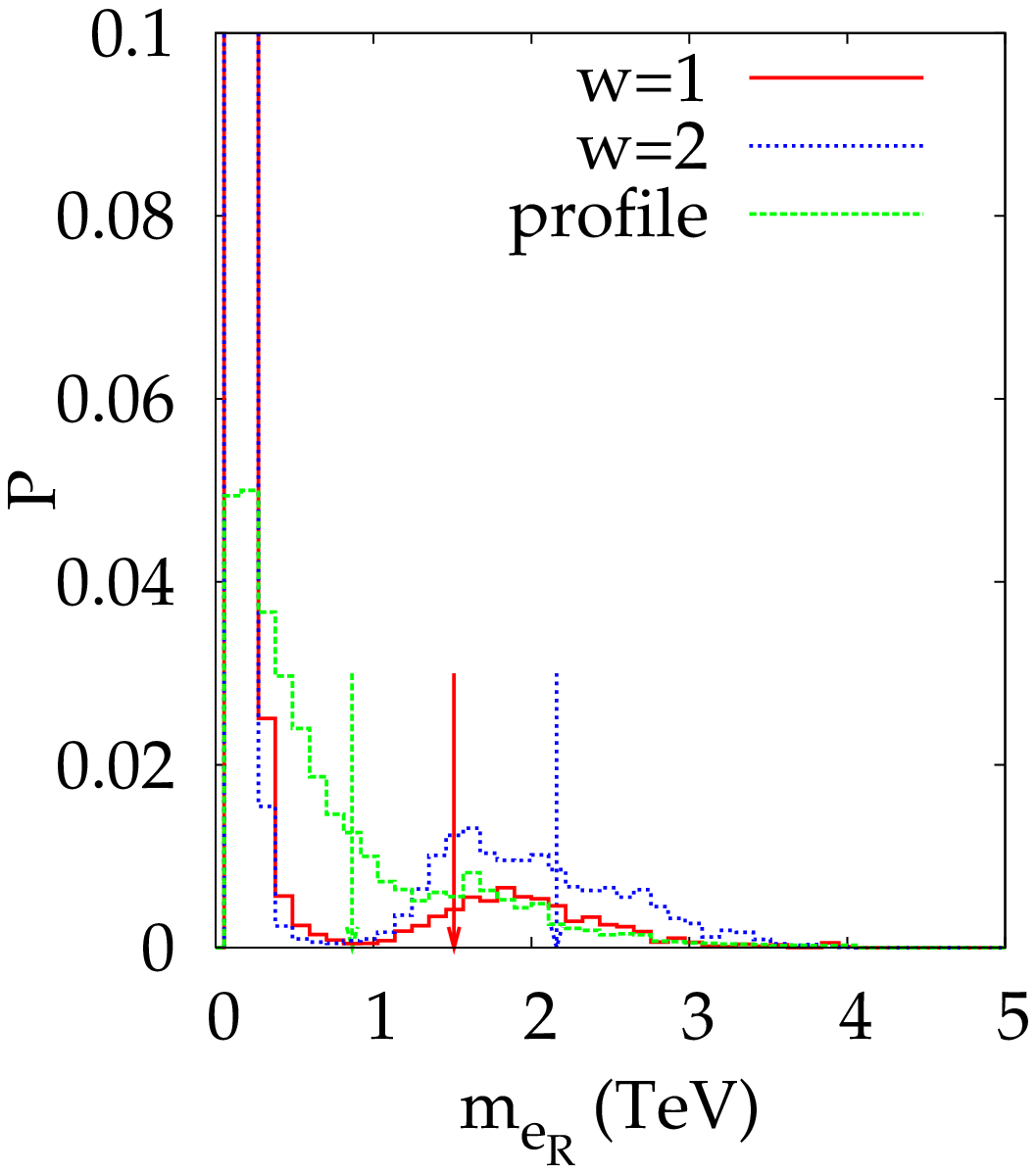}{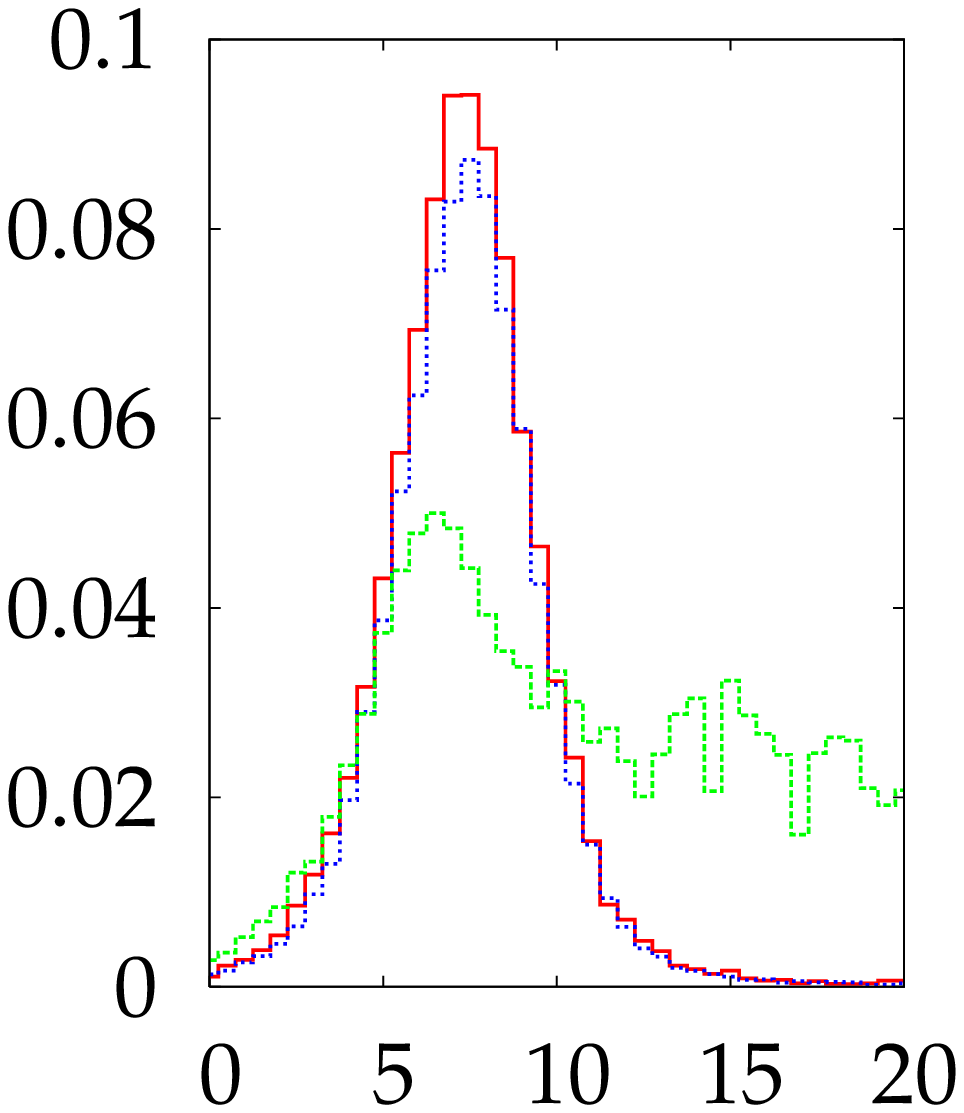}
\caption{MSSM particle mass pdfs and profile likelihoods: dependence upon the
  prior in the CMSSM\@. The 
  vertical arrows display the one-sided 95$\%$ upper limits on 
  each mass. There are 75 bins on each abscissa. Histograms marked ``profile''
  are discussed in section~\protect\ref{sec:profile} and have been multiplied
  by different dimensionful constants in order to be comparable by eye with the
  $w=1,2$ pdfs. The profile 95$\%$ confidence level upper limits are
  calculated by finding the 
  position for which the 1-dimensional profile likelihood has 
$2 \Delta \ln L = 2.71$~\cite{minuit}.  
\label{fig:masses}}
}
Pdfs of sparticle and Higgs masses coming from the fits are displayed in
  Figs.~\ref{fig:masses}a-\ref{fig:masses}h along with 95$\%$ upper bounds
  calculated from the   pdfs. 
The pdfs displayed are for the masses of (a) the lightest CP even Higgs, (b)
  the CP-odd Higgs, (c) the left-handed squark, (d) the gluino, (e) the
  lightest neutralino, (f) the lightest chargino, (g) the right-handed
  selectron and (h) the lightest-stau lightest-neutralino mass
  splitting respectively.
The most striking feature of the figure is that 
the Higgs and sparticle masses tend to be very light for the REWSB and same
  order prior, boding well for future collider sparticle searches. This effect
  is consistent with 
  a preference for smaller $m_0$, $M_{1/2}$ exhibited by the new priors in
  Fig.~\ref{fig:priors}b,d. In general, there is remarkably little difference
  between the two different cases of $w=1$ or $w=2$.
  This fact is perhaps not so surprising considering that the shape of the
  priors doesn't change enormously with $w$, as
  Figs.~\ref{fig:ft},\ref{fig:priors} show.   
  The sparticle
  mass   distributions for priors that are flat in $\tan \beta$ were displayed
  in Refs.~\cite{Allanach:2005kz,deAustri:2006pe,darkSide} and show a spread
  up to much higher values of the masses.  
  As we have explained above, we do not believe flat $\tan \beta$ to be an
  acceptable prior. Some readers may consider it to be so: such readers may
  consider our fits to be considerably less robust to changes in the prior
  than Fig.~\ref{fig:masses} indicates. 
  Lower values of $A_0$ and $\tan \beta$ help
  to make the lightest CP-even Higgs light in the REWSB+same order prior
  case, shown in Fig.~\ref{fig:masses}a. 
  The mass ordering $m_{{\tilde q}_l} > m_{\chi_2^0} > m_{{\tilde l}_R} >
  m_{\chi_1^0}$ allows a ``golden channel'' decay chain of ${\tilde q}_l
  \rightarrow  {\chi_2^0} 
  \rightarrow {\tilde l}_R \rightarrow m_{\chi_1^0}$. Such a decay chain has
  been used to provide several important and accurate constraints upon the
  mass spectrum~\cite{Allanach:2000kt}. In some regions of parameter space, it
  can also allow spin information on the sparticles involved to be
  extracted~\cite{Barr}. 
  We may calculate the Bayesian posterior probability
  of such circumstances by integrating the posterior pdf over the parameter
  space that allows such a mass ordering. From the MCMC this is simple: we
  simply count the fraction of sampled points that have such a mass
  ordering\footnote{Other absolute probabilities quoted below are calculated
  in an analogous manner.}. 
  The posterior probability of such a mass ordering is
  high: 0.93  for $w=1$ and 0.85 for $w=2$, indicating that analyses using the
  decay chain are
  likely to be possible (always assuming the CMSSM hypothesis, of course).

  As pointed out in 
  Ref.~\cite{Allanach:2005kz}, the flat $\tan \beta$ posteriors extend out to
  the 
  assumed upper range taken on $m_0$ and so the flat $\tan \beta$ pdf for the
  scalar masses were artificially cut off at the highest masses displayed. 
  This is no longer the case for the new choice of priors since the regions of
  large posterior do   not reach the chosen ranges of parameters, as shown in
  Figs.~\ref{fig:newprior}b,d. 
  Thus our derived upper bounds on, for instance $m_{{\tilde q}_L}$ in
  Fig.~\ref{fig:masses}c and $m_{{\tilde e}_R}$ in Fig.~\ref{fig:masses}g are
  not dependent upon the $m_0<4$ TeV range chosen.
  The mass splitting between the lightest stau and the neutralino is displayed
  in Fig.~\ref{fig:masses}h. The insert shows a blow-up of the
  quasi-degenerate stau-co-annihilation region and has a different
  normalisation to the rest of the plot. Since the REWSB+same order prior fit results lie
  in the co-annihilation region, nearly all of the probability density
  predicts that $m_{{\tilde \tau}_1} - m_{\chi_1^0}<20$ GeV. It is a subject
  of ongoing research as how to best verify this at the
  LHC~\cite{Arnowitt:2006xw}. 
  In Fig.~\ref{fig:masses}g, the plot has been cut off at a probability $P$ of
  0.1 and the histograms actually extend to 0.70,0.68
  in the lowest bin for $w=1$ and $w=2$ respectively. Similarly, we have cut
  off Fig.~\ref{fig:masses}h at a probability 
  of 0.05. The fits extend to 0.93, 0.85
  for $w=1$, $w=2$ respectively in the lowest bin.

\FIGURE{\sixgraphs{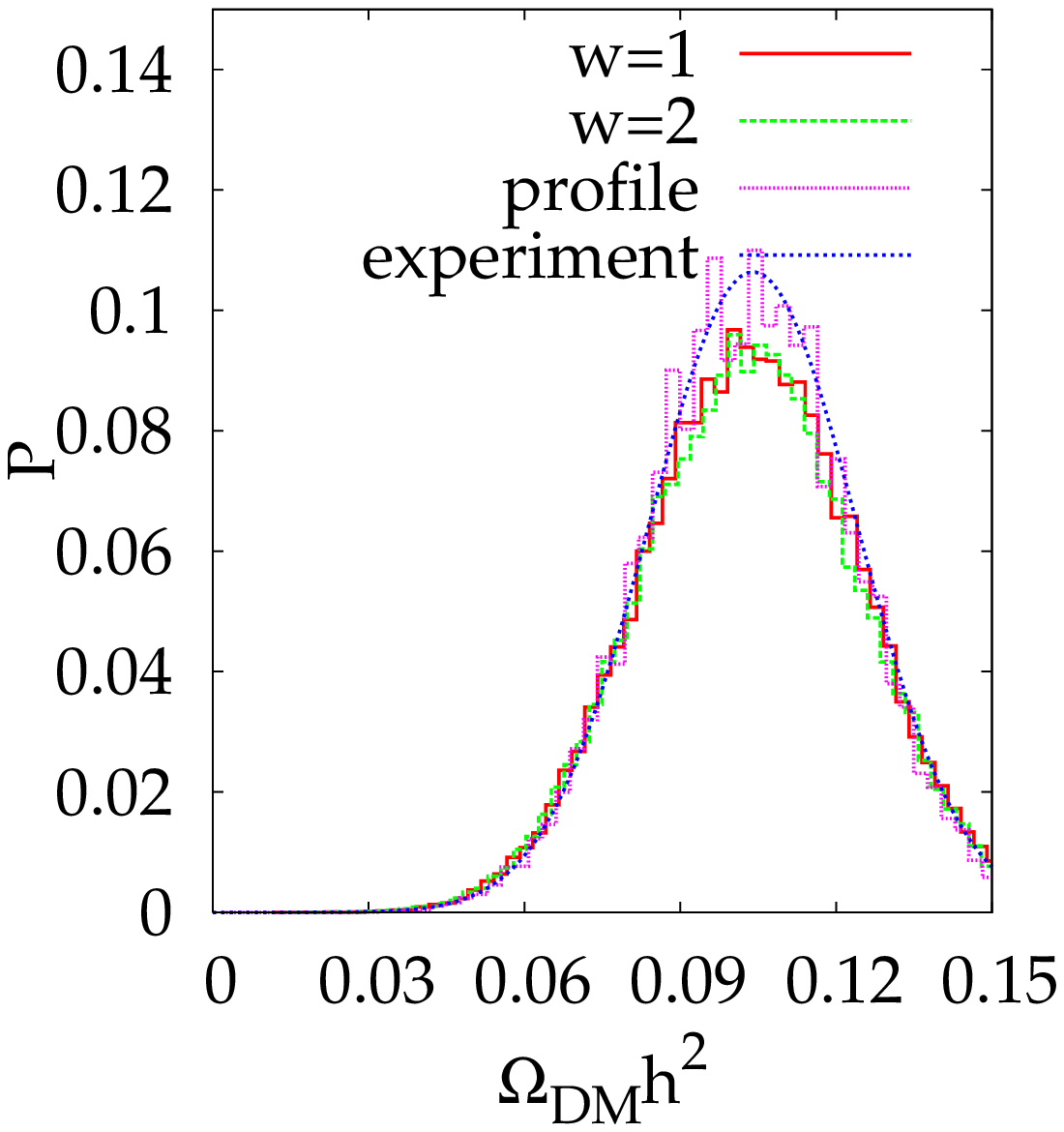}{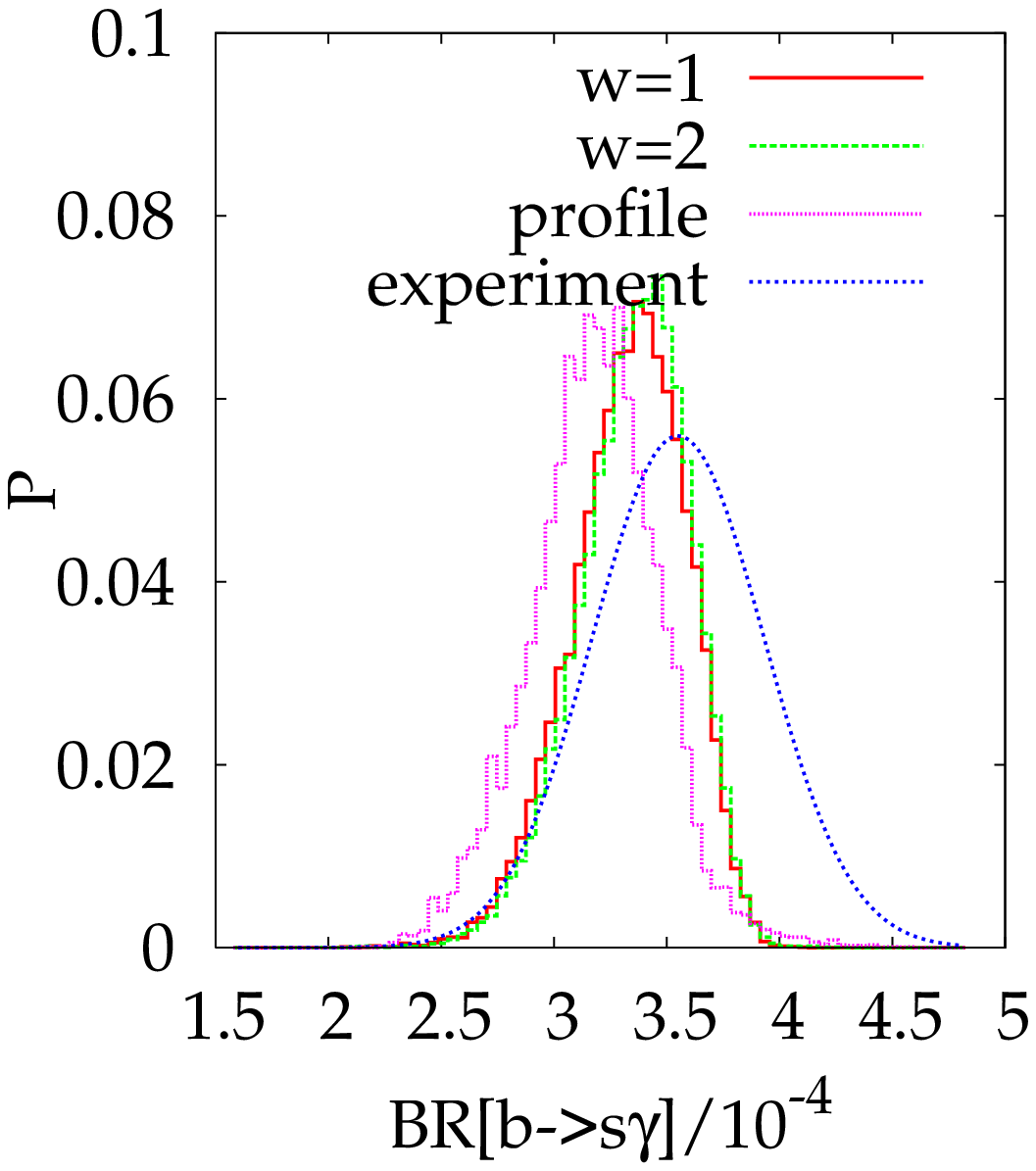}{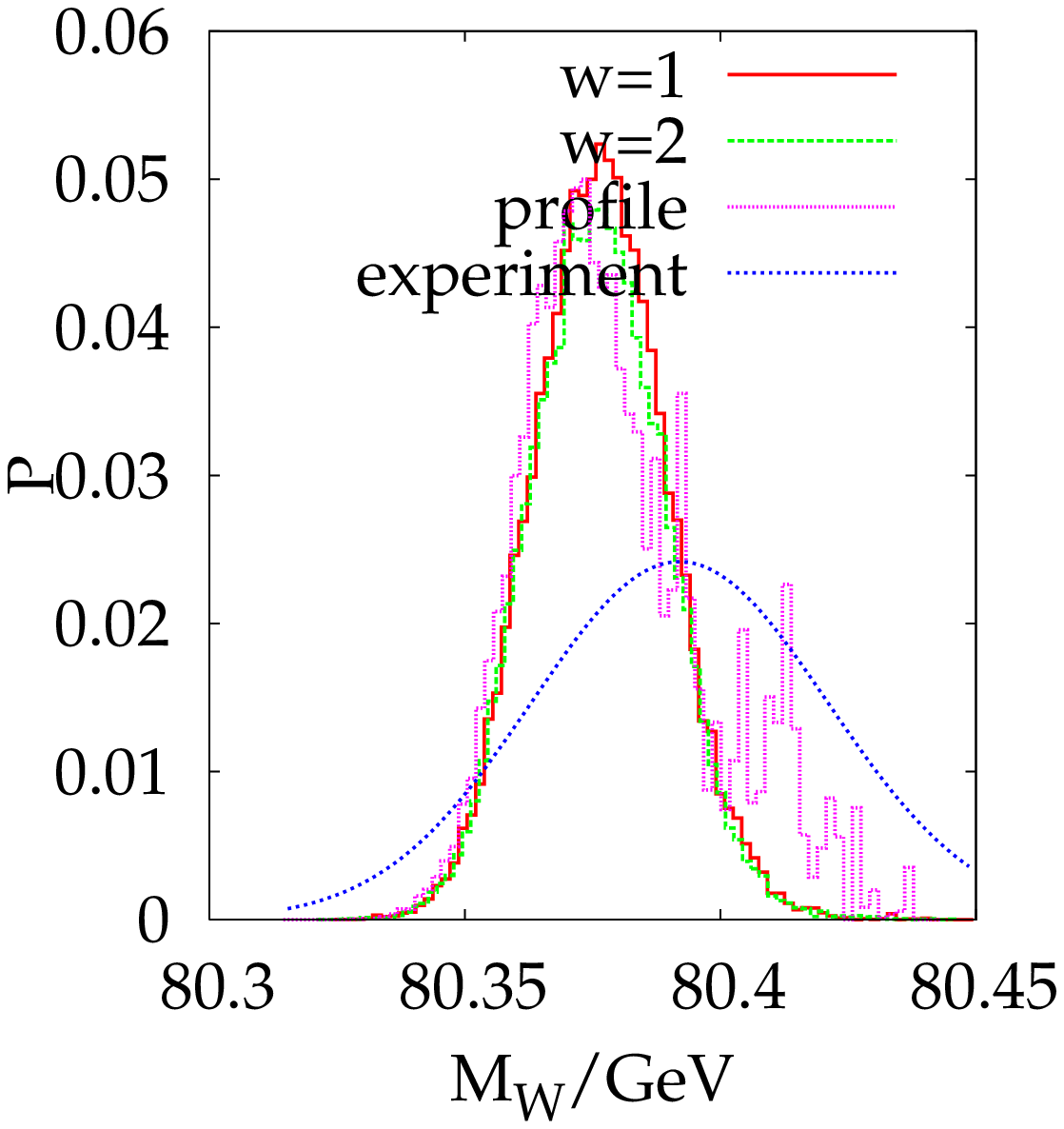}{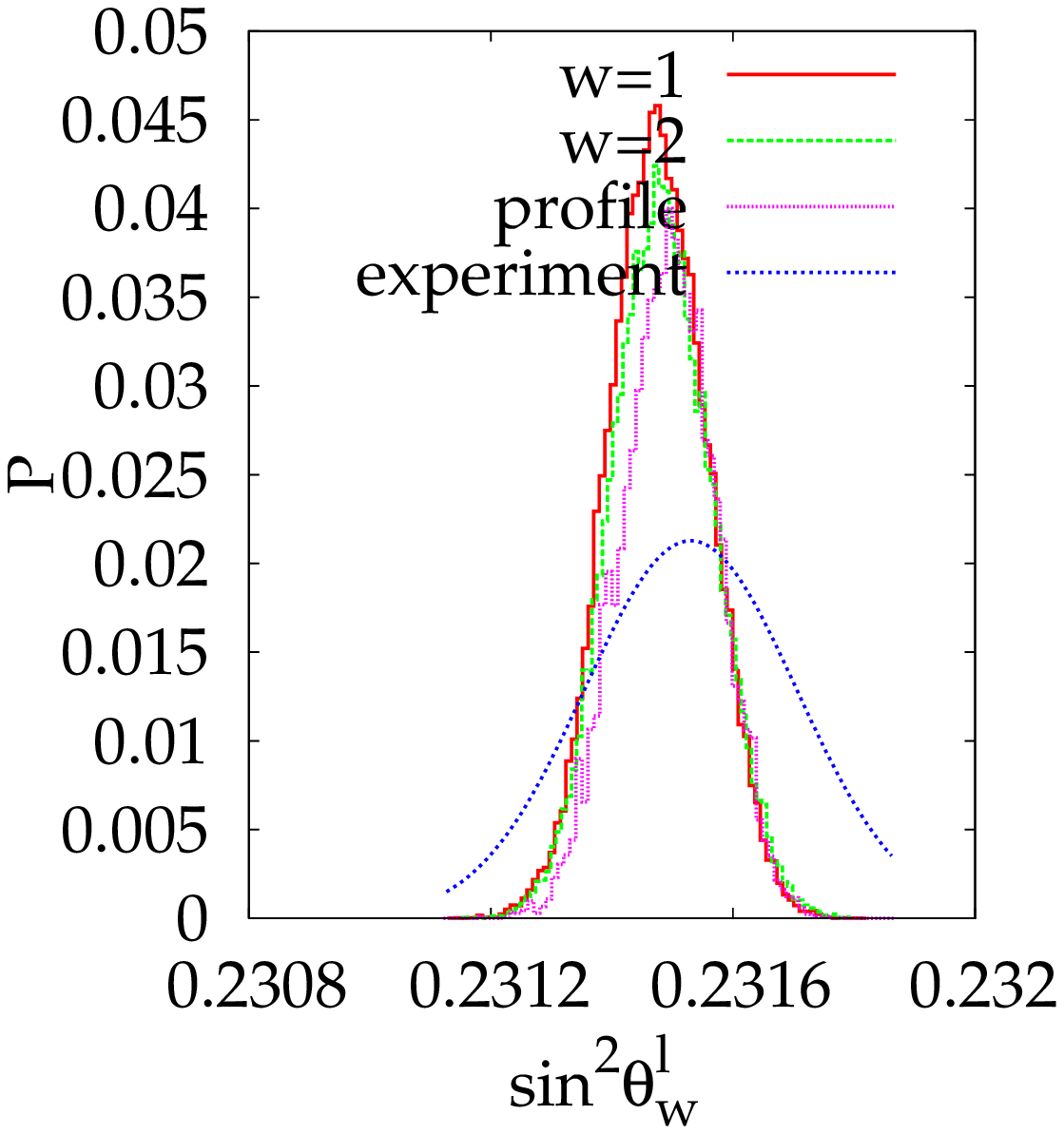}{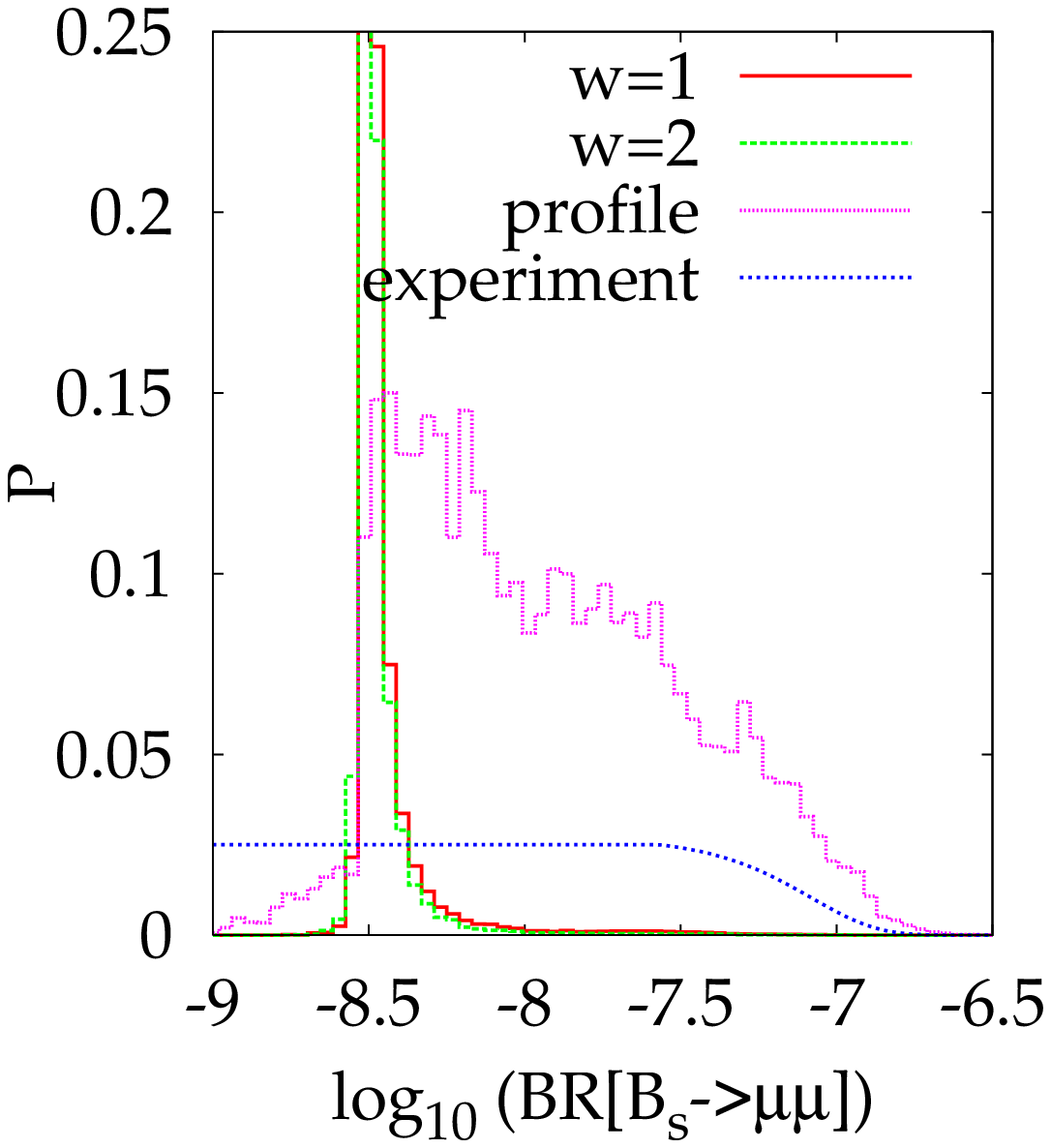}{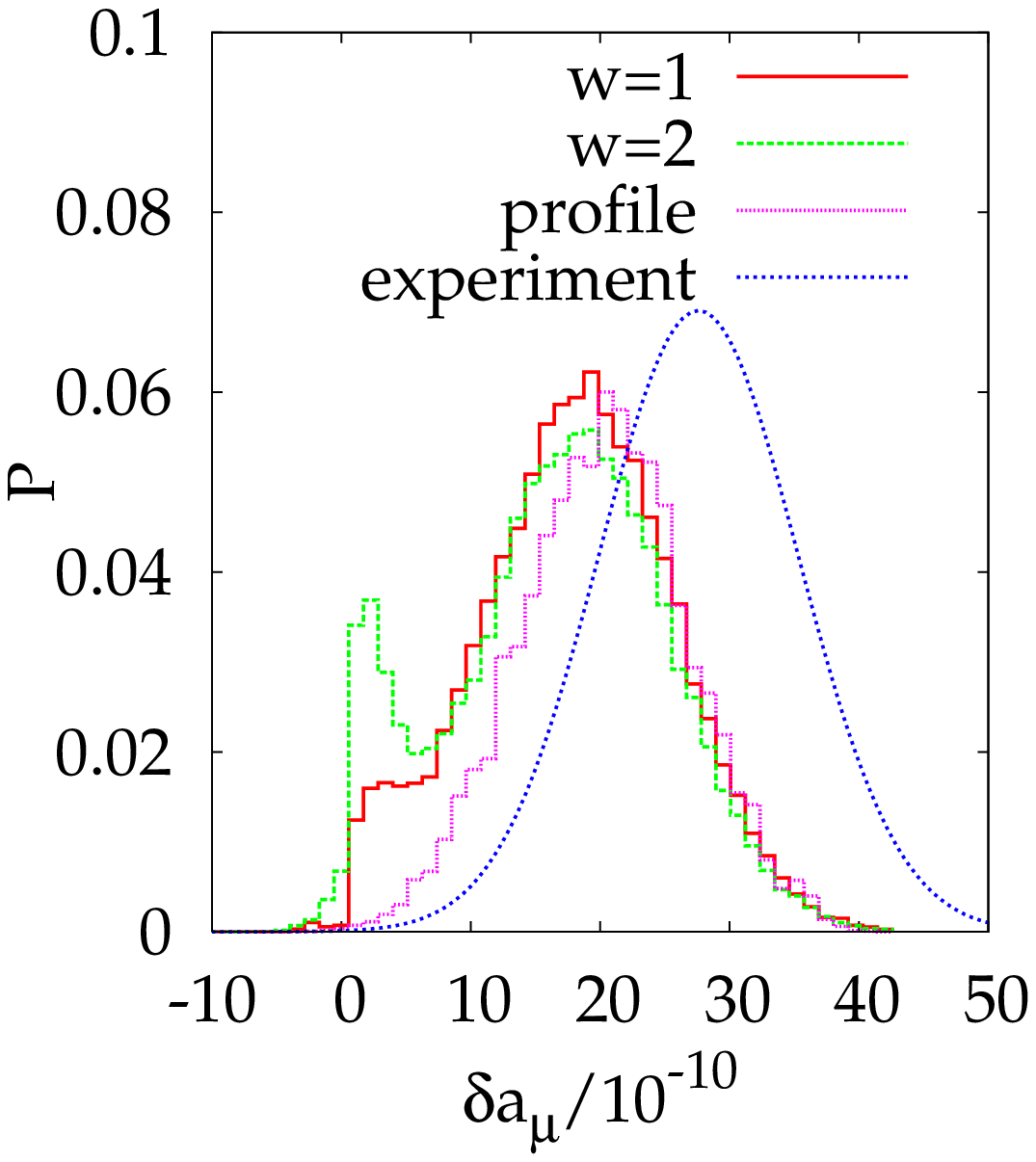}
\caption{Statistical pull of different observables in CMSSM fits. We show the
  pdfs for the experimental measurements as well as the posterior pdf of the
  predicted distribution in 
  $w=1$ and $w=2$ fits. Profile histograms
  are discussed in section~\protect\ref{sec:profile} and are multiplied
  by different dimensionful constants in order to be comparable by eye with the
  $w=1,2$ pdfs. 
\label{fig:obs}}}
We examine the statistical pull of the various observables in
Fig.~\ref{fig:obs}. In each case, the likelihood coming from the empirical
constraint is shown by the continuous distribution. The histograms show the
fitted posterior pdfs depending upon the prior. We have sometimes slightly
altered the normalisation of the curves and histograms to allow for clearer
viewing. 
Fig.~\ref{fig:obs}a shows that
the $\Omega_{DM} h^2$ pdf is reproduced well by all fits irrespective of which
prior distribution is used. This is because the fits are completely
dominated by the $\Omega_{DM} h^2$ contribution, since the CMSSM parameter space
typically predicts a much larger value than that observed by
WMAP~\cite{darkSide}. 
Figs.~\ref{fig:obs}b,\ref{fig:obs}c,\ref{fig:obs}d show that 
$BR[b \rightarrow s \gamma]$, $M_W$, $\sin^2 \theta_w^l$ are all constrained
to be near their central values, with less variance than is required by the
empirical constraint. Direct sparticle search limits mean that sparticles
cannot be too light and hence cannot contribute strongly to the three
observables. 
The rare decay branching ratio
$BR[B_s\rightarrow \mu\mu]$ is displayed in Fig.~\ref{fig:obs}e. Both fits
are heavily peaked  
around the SM value of $10^{-8.5}$, indeed the most probable bin has been
decapitated in the figure for the purposes of clarity, and really should
extend up to a probability of around 0.9. The SUSY
contribution to $BR(B_s\rightarrow
\mu\mu) \propto \tan \beta^6 / M^4_{SUSY}$ and so the preference for small
$\tan \beta$ beats the preference for smallish sparticle masses $\sim
O(M_{SUSY})$ in the new fits. In all of Figs.~\ref{fig:obs}a-e,
changing the width of the priors from 1 to 2 has negligible effect on the
results. The exception to this trend is $\delta a_\mu$, as shown in
Fig.~\ref{fig:obs}f. $\delta a_\mu$ has a shoulder around zero for $w=2$,
corresponding to a small amount of posterior probability density at high
scalar masses, clearly visible from Fig.~\ref{fig:masses}g. Such high masses 
suppress loops responsible for the SUSY contribution to $(g-2)_\mu$. 
$\delta a_\mu$ is pulled to lower values than the
empirically central value by direct sparticle limits and the preference for
values of $\tan \beta$ that are not too large. The almost negligible portion
of the graph for which $\delta a_\mu<0$ corresponds to $\mu<0$ in
the CMSSM. $(g-2)_\mu$ has severely suppressed the likelihood, and therefore
the posterior, in this portion of parameter space. For flat $\tan \beta$
priors, and $\delta a_\mu=22 \pm 10 \times 10^{-10}$, we had previously
estimated  
that the ratio of integrated posterior pdfs between $\mu<0$ and $\mu>0$ 
was $0.7-0.16$. For the new priors, where sparticles are forced to be
lighter, their larger contribution to $\delta a_\mu$ further suppresses the
$\mu<0$ posterior pdf. From the samples, we estimate\footnote{These numbers come from the
  mean and standard deviation of 10 chains, each of which is considered to
  deliver an independent estimate.} $P(\mu<0) /
P(\mu>0)=0.001 \pm 002$ for $w=1$ and $0.003\pm0.003$ for $w=2$, respectively for $\delta
a_\mu=(27.6\pm7.7) \times 10^{-10}$. Thus, while the probabilities are
not accurately determined, we know that they are small enough to neglect the
possibility of $\mu<0$. 

\section{Profile Likelihoods\label{sec:profile}}

Since, for a flat prior, Eq.~\ref{bayes} implies that the posterior is
proportional 
to the likelihood in a Bayesian analysis, one can view the distributions
resulting 
from the MCMC scan as being a ``likelihood map''~\cite{Allanach:2005kz}. 
If one marginalises in the unseen dimensions in order to produce a one or
two-dimensional plot, one either interprets the resulting distribution
probabilistically in terms of the posterior, or alternatively as a way of
viewing the full $n$-dimensional likelihood map, but without a probabilistic
interpretation in terms of confidence limits, or credible intervals. 
Instead, frequentist often eliminate unwanted parameters (nuisance parameters)
by maximization instead of marginalization.  The likelihood function of the
reduced set of parameters with the unwanted parameters at their conditional
maximum 
likelihood estimates is called the profile likelihood~\cite{profile}.
Approximate confidence limits can be set by finding contours of likelihood
that differ from the best-fit likelihood by some amount. This amount depends
upon the number of ``seen dimensions'' and the confidence level, just as in a
standard $\chi^2$ fit~\cite{minuit}. 

While we believe that dependence on priors actually tells us something useful
about the robustness of the fit, we are also aware that many high energy
physicists find the dependence upon a subjective measure distasteful, and
would be happier with a frequentist interpretation. When the fits are robust,
i.e.\ there is plentiful accurate data, we expect the Bayesian and frequentist
methods to identify similar regions of parameter space in any fits. We are not
in such a situation with our CMSSM fits, as we have shown in previous
sections, and so we provide the profile likelihood here for completeness. 

We can use the scanned information from the MCMC chains to extract the
profile likelihood very easily. Let us suppose, for instance, that we wish to
extract the profile in $m_0-M_{1/2}$ space. We therefore bin the chains
obtained in $m_0-M_{1/2}$ as before. We find the maximum likelihood in the
chain for each bin and simply plot that. The 95$\%$ confidence level region
then is delimited by the likelihood contour at a value $2 \Delta \ln L =
5.99$~\cite{minuit}, where $\Delta \ln L = \ln L_{max} - \ln L$. 
The profile likelihoods in the $m_0-M_{1/2}$ and $m_0-\tan \beta$ plane are
shown in Fig.~\ref{fig:2dprofiles}. 
\FIGURE{\twographs{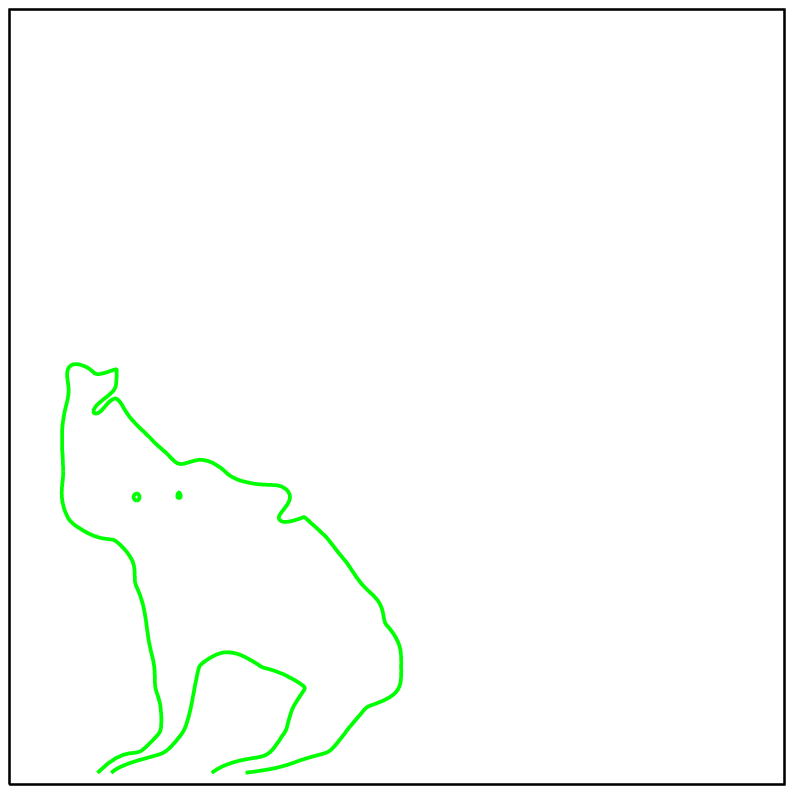}{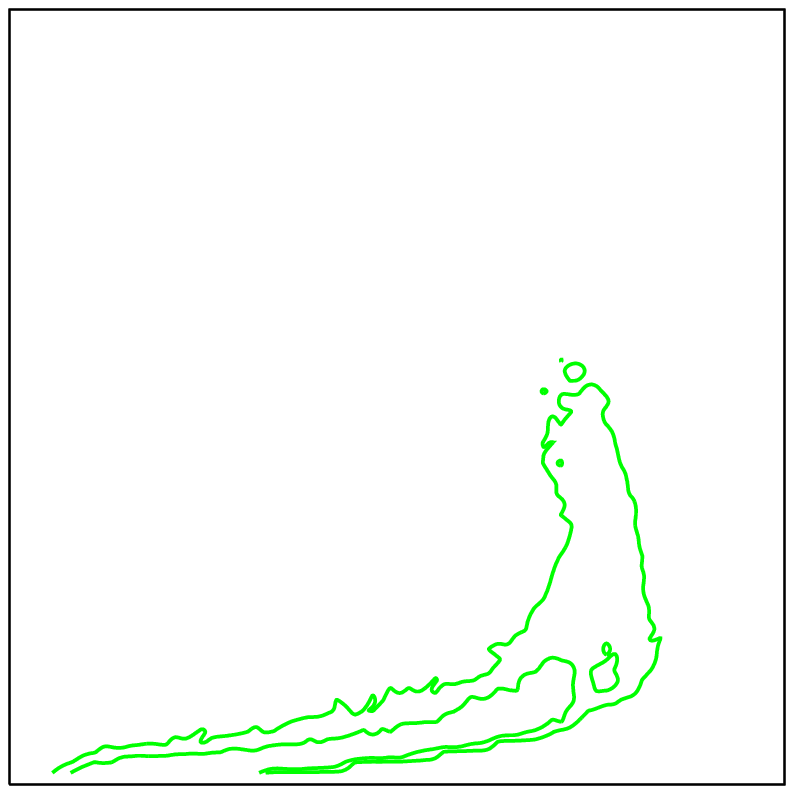}
\caption{Two dimensional profile likelihoods in the 
(a) $m_0-M_{1/2}$ plane, (b) $m_0-\tan \beta$ plane. There are 75 bins along
  each direction. The inner   (outer) contours
  show the 68$\%$ and 95$\%$ confidence level regions
  respectively.\label{fig:2dprofiles}}}  
Comparing Figs.~\ref{fig:2dprofiles}a and \ref{fig:newprior}a, we see that the profile likelihood gives
similar information to the Bayesian analysis with flat likelihoods. The main
difference is that the profile likelihood's confidence limit only extends out
to $(M_{1/2}, m_0) < (1.0, 2)$ TeV, whereas for the Bayesian flat-prior
analysis, values up to 
$(M_{1/2}, m_0) < (1.5, 4)$ TeV are viable.
Comparing Fig.~\ref{fig:2dprofiles}b and \ref{fig:newprior}c, we again see 
similar constraints, except that the tail at high $\tan \beta$ up to larger
values of $m_0>2$ TeV has been suppressed in the profile. From the difference
we learn the following facts: in this high $\tan \beta$-high $m_0$ tail, the
fit to data is less good than in other regions of parameter space. However, it
has a relatively large volume in unseen dimensions of parameter space, which
enhances the posterior probability in Fig.~\ref{fig:newprior}c. 
The difference between the two plots is therefore a good measure of such a 
so-called ``volume effect''. In ref.~\cite{deAustri:2006pe,rosz2},
an average-$\chi^2$ estimate was 
constructed in order to identify such effects. We find the profile likelihood
to be easier to interpret, however. It also has the added bonus of allowing a
frequentist interpretation.

We show the profile likelihoods of the various relevant masses in
Fig.~\ref{fig:masses}.  There is a general tendency for all of the masses
to spread to somewhat heavier values than the $w=1,2$ same order+REWSB
priors. 
We remind the reader that the profile likelihood histograms are not pdfs. In
the figure, they have been multiplied by dimensionful constants that make them
comparable eye to the Bayesian posteriors on the plot.
The gluino mass shows the most
marked difference: it appears that higher gluino masses are disfavoured by 
volume effects in the Bayesian analyses. However, while the profiles differ
from the Bayesian analyses to a much larger degree than the $w=1$ or $w=2$
prior fits differ from each other, 
they are not wildly different to the Bayesian analyses. The higgs mass
distributions look particularly similar. There is a qualitative difference in
Fig.~\ref{fig:masses}g,h, where $m_{{\tilde e}_R}$ and $m_{{\tilde \tau}_1} -
m_{\chi_1^0}$ have a 
non-negligible likelihood up to 1 TeV, unlike the
posterior probabilities. 

Figs.~\ref{fig:obs}a-f show the profile likelihoods of the pull of various
observables. We see that $\Omega_{DM} h^2$ shows a negligible difference to the
posteriors. This is because the dark matter relic density constraint
dominates the fit and determines the shape and volume of the viable parameter
space. Most of the profiles are similar to the posteriors in the figure except
for Fig.~\ref{fig:obs}e, where the likelihood extends out to much higher
values of the branching ratio of $B_s \rightarrow \mu \mu$. These values
correspond in Fig.~\ref{fig:2dprofiles}b to high $\tan \beta$ but low $m_0$
points. The posteriors for high $BR(B_s \rightarrow \mu \mu)\propto
1/{M_{SUSY}}^2$ are suppressed 
because of the large volumes at high $m_0$ (and hence at high $M_{SUSY}$,
where $BR(B_s \rightarrow \mu \mu)$ approaches the Standard Model limit due to decoupling). 
In
Fig.~\ref{fig:obs}c, we see enhanced statistical fluctuations in the upper
tail of the profile likelihood of $M_W$, presumably due to a small number of
sampled points there. These fluctuations could be reduced with further running
of the MCMCs, however.

\section{LHC SUSY Cross Sections \label{sec:crossec}}

In order to calculate pdfs for the expected CMSSM SUSY production
cross-sections at 
the LHC, we use {\tt HERWIG6.500}~\cite{herwig} with the default parton
distribution functions. We calculate the total cross-section of the production
of two sparticles with transverse momentum $p_T>100$ GeV.
We take the fitted probability distributions of the
previous section with the REWSB+same order priors and use {\tt   HERWIG6.500} to calculate cross-sections for 
(a) strong SUSY production i.e.\ squark and gluino production,
(b) inclusive weak gaugino production (i.e.\ a neutralino or chargino in
association with another neutralino, a chargino, a gluino, a squark or a
gluino) and (c) 2-slepton production. No attempt is made here to fold in
experimental efficiencies or the branching ratios which follow the decays into
final state products. The total cross-section times assumed integrated
luminosity therefore serves as an upper-bound on the number of events expected
at the LHC in the different channels (a)-(c). Some analyses
give a few percent for efficiencies, but for specific cases of more
difficult signatures, the efficiencies can be tiny.

We show the one dimensional pdfs for the various SUSY production
cross-sections in Fig.~\ref{fig:sigmas}a. We should bear in mind that the LHC
is expected to deliver 10 fb$^{-1}$ of luminosity per year in
``low-luminosity'' mode, whereas afterward this will increase to 30
fb$^{-1}$. Several years running at $\log_{10} \sigma/$fb$=0$ therefore
corresponds 
to of order a hundred production events for 100 fb$^{-1}$. $\log_{10}
\sigma/$fb$=0$ then gives some kind of rough limit for what might be observable
at the LHC, once experimental efficiencies and acceptances are factored in.
Luckily, we see that strong production and inclusive weak gaugino production 
are always above this limit, providing the optimistic conclusion that SUSY
will be discovered at the LHC (provided, as always in the present paper, that
the CMSSM hypothesis is correct and that the reader accepts our proposal for
the prior pdfs). The 95$\%$ lower limits on the
total direct production cross-sections are {360} fb, {90} fb and {0.01} fb for    
strongly interacting sparticle, inclusive weak gaugino and slepton
production respectively.  There therefore is a small chance that direct slepton
production  may not be at observable rates. 
The posterior probability that $\sigma(p p \rightarrow \tilde l^+ \tilde
l^-)<1$ fb is 0.063. 
Even in the event that direct slepton production is at too slow a rate to be
observable, it is possible that sleptons can be observed and measured by the
decays of other particles into them~\cite{Allanach:2000kt}.
The pdfs of total SUSY production cross-sections for $w=2$ are almost
identical to those shown in the figure. The main difference is in the total
direct slepton production cross section, where the small bump at $\sigma \sim
10^{-2}$ fb is somewhat enlarged. It has the effect of placing the 95$\%$
lower bound on the slepton production cross-section 
at 4.8$\times 10^{-4}$ fb. For $w=2$, the chance of the di-slepton production
cross-section being less than 1 fb is 0.15.
The strong and weak gaugino production
cross-sections have 95$\%$ lower bounds of 570,90 fb respectively for $w=2$.
\FIGURE{\fourgraphstt{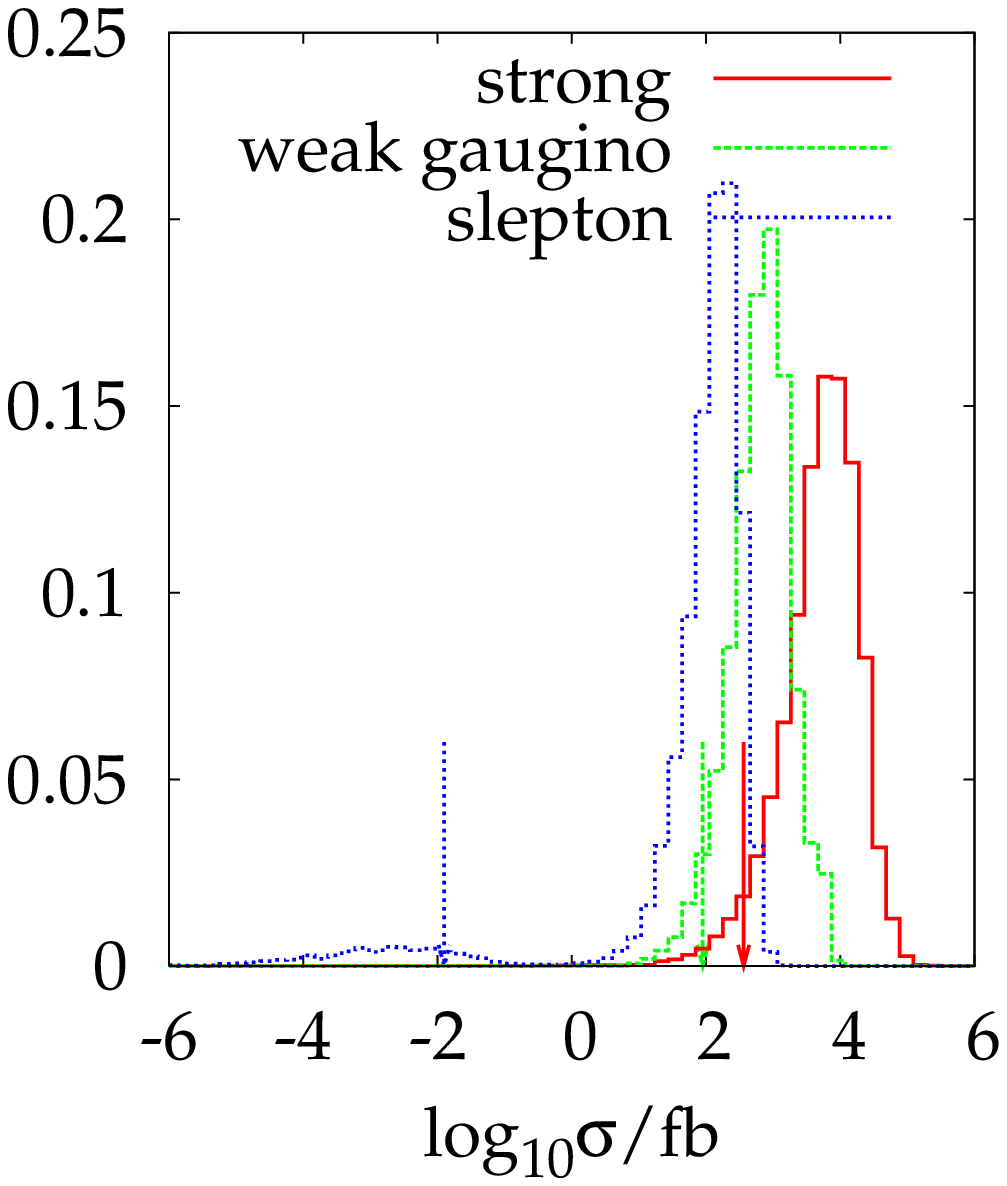}{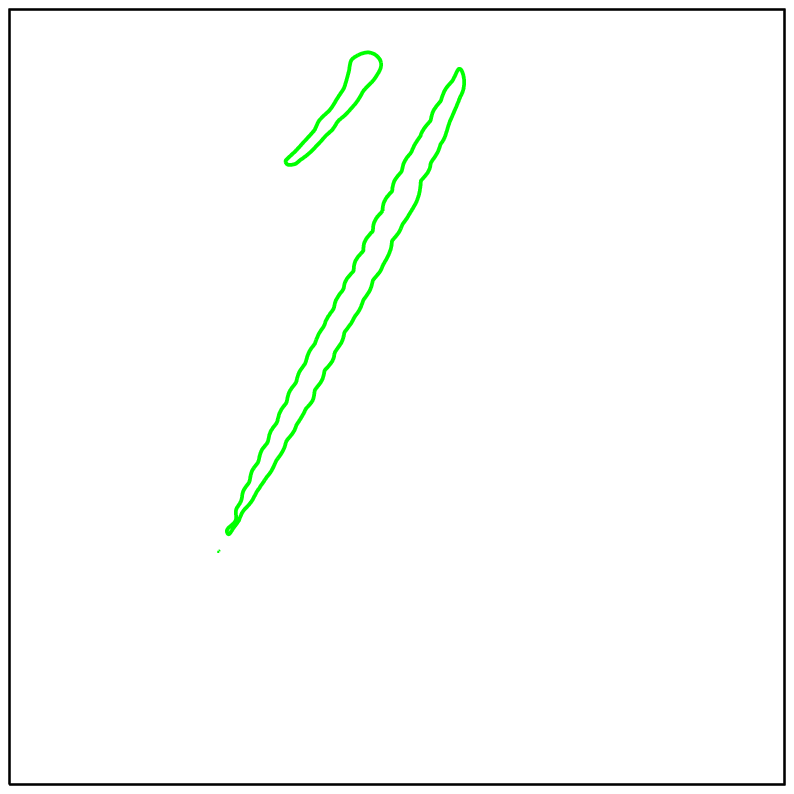}{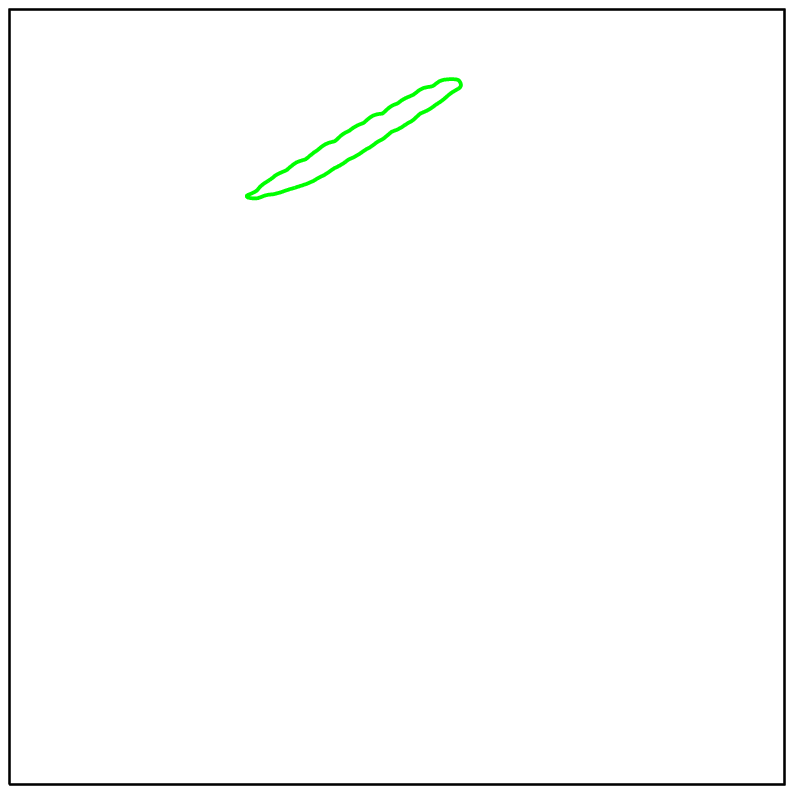}{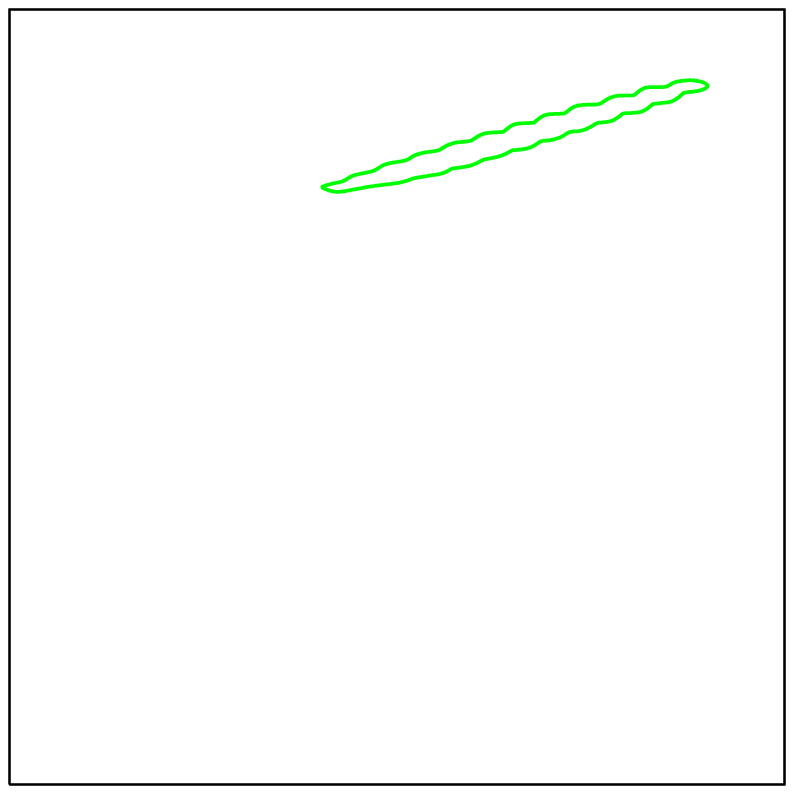}
\caption{Total SUSY LHC production cross-section pdfs in the CMSSM 
  with REWSB+same order $w=1$ priors. ``strong'' refers to squark/gluino
  production, ``weak'' to 
  inclusive weak gaugino production and ``slepton'' to direct slepton
  production. In (a), 95$\%$ {\protect\em lower} \/limits on the cross-sections
  are shown by the   vertical arrows.
  The probability normalised to the bin with maximum probability,
  is shown by reference to the colour-bar on the right hand side for (b), (c)
  and (d). The contours show the 95$\%$ limits in the two-dimensional plane. 
  \label{fig:sigmas}
}}

We examine correlations between the various different cross-sections in
Figs.~\ref{fig:sigmas}b-d. For instance, Fig.~\ref{fig:sigmas}b has two
distinct maxima, the focus-point region on the left-hand side and the stau
co-annihilation region on the right-hand side. If one could obtain empirical
estimates of the total cross-sections to within a factor of about 3
(corresponding to an error of about 0.5 in the $\log_{10}$ value)
then measurements of
$\sigma_{\mathrm{strong}}$ and $\sigma_{\mathrm{weak}}$ 
could distinguish between the two mechanisms.
There is a overlap between the one-dimensional projections
of the two different regions in either $\sigma_{\mathrm{strong}}$ 
or $\sigma_{\mathrm{weak}}$ and so measurements of both seem to be required for
discrimination. The
probability density of the focus-point region becomes too smeared in the
$\sigma_{\mathrm{slepton}}$ direction to appear in the 95$\%$ limit bounds in
Fig.~\ref{fig:sigmas}c,d. Experimental measurements of the
cross-sections in Fig.~\ref{fig:sigmas} would provide a test of the CMSSM
hypothesis. 
It is clear from Fig.~\ref{fig:sigmas}a that
$\sigma_{\mathrm{slepton}}$ has two isolated probability maxima. The one at
$\sigma_{\mathrm{slepton}}<0$ corresponds to the focus point region, where
scalar x
masses are large. This region will probably directly produce too few sleptons
to be observed at the LHC and so will not be useful there for discriminating
the CMSSM focus point region from the co-annihilation region unless there is a
significant luminosity upgrade~\cite{slhc}. 

\FIGURE[r]{\unitlength=1.1in
\begin{picture}(2.2,2.5)(0,0)
\put(-0.2,0){\epsfig{file=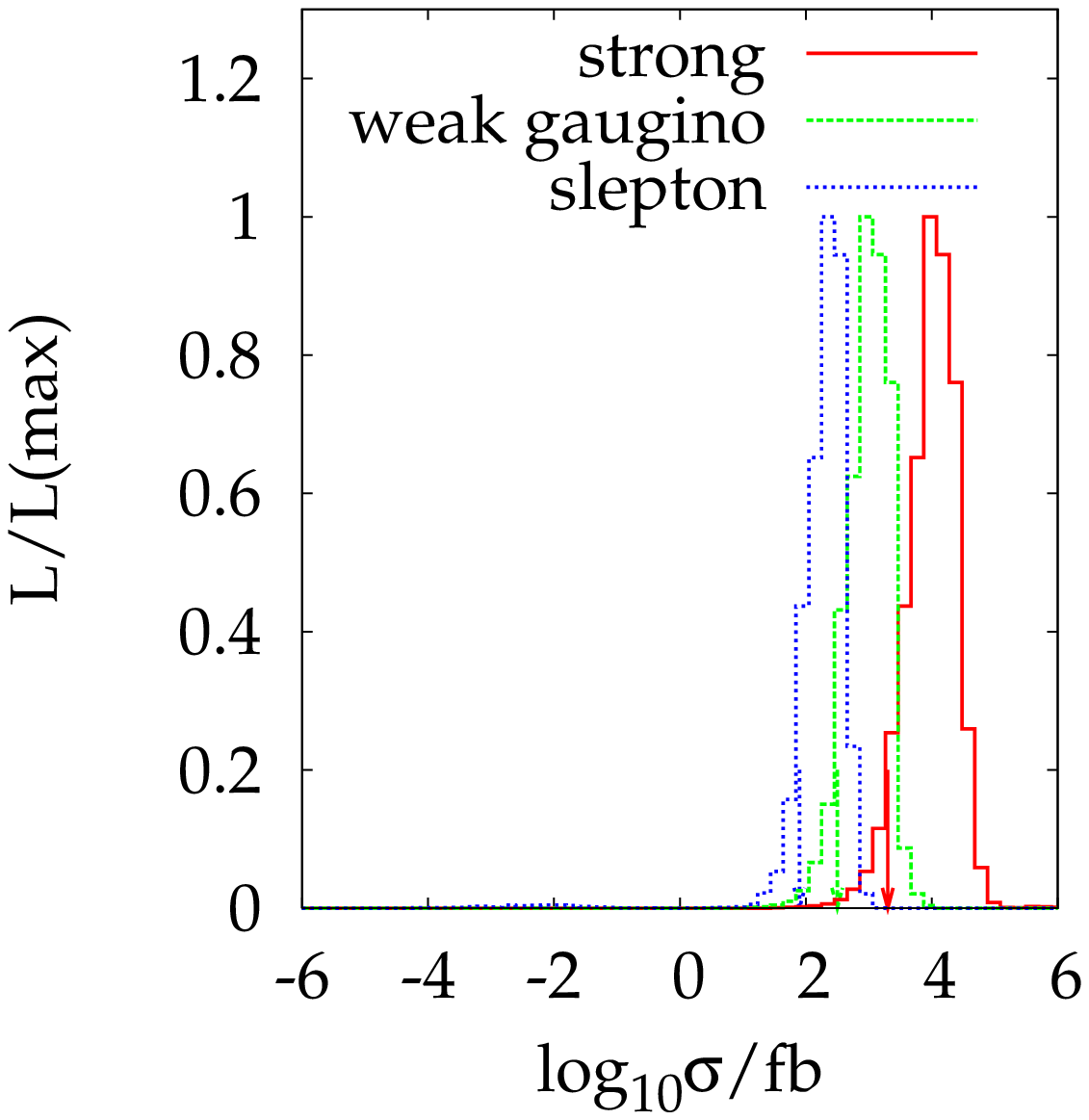, width=2.9in}}
\end{picture}
\caption{SUSY production cross-section profile likelihoods. One-sided 95$\%$
  lower confidence 
  level limits are shown as calculated from these histograms by the vertical
  arrows. 
  \label{fig:sigprof}}} 
The profile likelihoods of SUSY production cross-sections are shown in
Fig.~\ref{fig:sigprof}. In the figure, ``strong'' refers to squark/gluino
  production, ``weak'' to 
  inclusive weak gaugino production and ``slepton'' to direct slepton
  production. By comparison to fig.~\ref{fig:sigmas}a, we see that the profile
  likelihoods generally prefer somewhat larger SUSY production cross-sections 
  than the Bayesian analysis with REWSB+same order $w=1$ priors. The 95$\%$
  one-sided lower confidence level bounds upon them are for 2000 fb for sparton
  production, 300 fb for weak gaugino production and 80 fb for slepton
  production. This last bound is particularly different from the Bayesian
  analysis since there the small probability for the focus-point r\'{e}gime,
  evidenced by the low bump to the left hand side of
  Fig.~\ref{fig:sigmas}a, was only pushed just above an integrated posterior
  pdfs of $5\%$   by volume effects.  

\section{Conclusion \label{sec:conc}}
This analysis constitutes the first use in a serious physics context of a new
``banked'' MCMC proposal function~\cite{bank}. 
This new proposal function has allowed us to sample simultaneously,
efficiently and correctly from both signs of $\mu$. 
The resulting sampling passed convergence tests
and therefore gave reliable estimates of LHC SUSY cross-section pdfs. 
MCMCs have also been used to determine the impact of potential future 
collider data upon the MSSM~\cite{Lester:2005je,Baltz:2006fm,rosz2}. 
The development of tools such as the banked proposal MCMC constitutes
a goal at least as important as the interesting physics results derived here.
In case they may be of use for future work,
we have placed the samples obtained by the banked MCMC on the internet, with
instructions on how to read them, at the following URL:
\begin{verbatim}
http://users.hepforge.org/~allanach/benchmarks/kismet.html
\end{verbatim}

We argued that prior probability distributions that are flat in $\tan
\beta$ are less natural than those that are flat in the more
fundamental Higgs potential parameters $\mu$, $B$ of the MSSM\@. We have
derived a more natural prior distribution in the form of
Eq.~\ref{priorsummary}, which is originally flat in $\mu$, $B$ and
also encodes our prejudice that $\mu$ and the SUSY breaking parameters
are ``of the same order''. There is actually a marginalisation over a
family of priors, and as such our analysis uses a hierarchical Bayesian prior
distribution. It should be noted that this prior pdf can replace
definitions of fine-tuning in the MSSM Higgs sector. Its use in
Bayesian statistics is well-defined, and we have examined its effect
on Bayesian CMSSM analysis. The main effect is to strongly disfavour the
Higgs-pole and focus point dark matter annihilation regions of CMSSM
parameter space.  The sparticle masses are then predicted to be
probably lighter than previously thought as a result of the new prior.
There is little difference in the results when one changes the widths of the
same order pdfs, but the results are very different to previous ones in the
literature where flat priors in $\tan \beta$ were examined. If one rejects the
prior flat in the SUSY breaking parameters, as we have advocated here, our
results appear rather 
robust with respect to changes in the prior. However, for readers that find
the same order priors too strong, one can view the difference between the flat
prior results and those using the same order priors as a result of uncertainty 
originating from scarce data. 
This dependence
upon priors does indicate the need for caution when interpreting our results; 
constraining data are currently too scarce to render the posterior pdfs
approximately independent of the prior assumption. 
We feel that the sensitivity to priors must be studied, and find the 
large dependence on priors consistent with something that is intuitively
obvious~\cite{goldstein};  
that a few pieces of indirect data are not sufficient to robustly constrain a
complex  
model of 8 parameters.  The frequentist analysis does not depend on any prior,
but it  
also does not allow us to inject reasonable assumptions about the naturalness
of the theory.  
A comparison between the likelihood profile and posteriors is ideal because it
contains  
information about volume effects in the Bayesian analyses. The frequentist
confidence levels 
on MSSM particle masses are different to Bayesian credible
intervals,  but within the same ball-park as each other. Thus we may infer
some rough limits, but to be conservative one might take the {\em least
  constraining} upper bound by any of the different methods. 
The lighter sparticles from the new priors result in more
optimistic total SUSY cross-section predictions for the LHC\@.
It would be interesting to see the footprints of other SUSY
breaking models to see whether the correlations between different
cross-sections  are a good discriminator~\cite{footprint}. 
\appendix

\section{Comparison With Previous Literature \label{app:comp}}
The flat-prior results may at first sight seem to be in contradiction with the
analysis of 
Ellis {\em et al}~\cite{Ellis:2004tc}, where a preference for light SUSY was
found  
from quite similar global fits to those in the present paper. They also fit
$M_W, \sin^2 \theta_w^l(\mbox{eff})$ as well as $(g-2)_\mu$, while using the
relic density of dark matter as a constraint.
In their paper, Ellis {\em et al} \/fixed $\tan 
\beta$, and all Standard Model inputs at their central experimental
values. 
For every value of $M_{1/2}$, $A_0$ scanned, $m_0$ is adjusted until the
central WMAP3 value of $\Omega_{DM} h^2$ results. 
The smearing due to the
finite error on $\Omega_{DM} h^2$ is very small and so it is argued that 
this procedure well approximates the full constraints upon parameter space.
We display the resulting constraint on the $A_0-M_{1/2}$ plane for $\tan
\beta=10$ and $\mu>0$ in Fig.~\ref{fig:elliscomp}a. The partial ellipses show
the authors' claimed 68$\%$ and 90$\%$ confidence level limits calculated with
$\Delta \chi^2=2.30,4.61$~\cite{Ellis:2004tc} 
from the best-fit point, marked by a cross. Actually,
since the confidence 
level regions are constrained within a
wedge-shape in the figure, the 68$\%$ (90$\%$) limits should not necessarily
correspond 
to $\Delta \chi^2=2.30(4.61)$ respectively. The regions shown on the figure
should therefore be re-calculated, by calculating what sort of probability
distribution $\Delta \chi^2$ has when trapped in such a wedge. 
\FIGURE{\twographst{ellis}{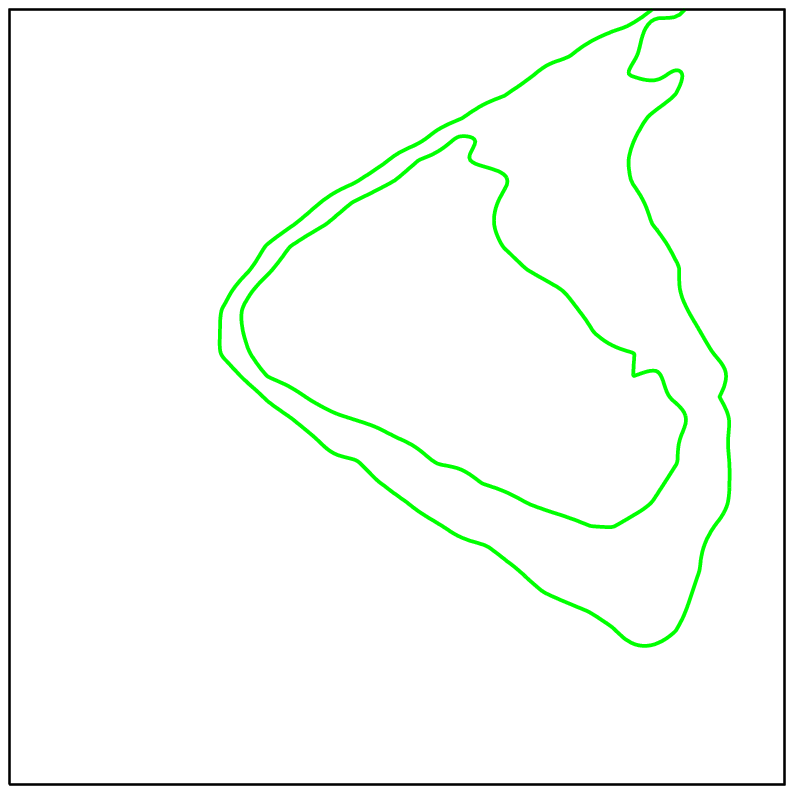}
\caption{(a) Reduced parameter space global fit from Ref.~\cite{Ellis:2004tc}
  for $\tan \beta=10$, $\mu>0$. In the plot, $A_0$ has a relative minus sign
  with respect to the definition used in the present paper, (b)
our version of the same fit, marginalised over $m_0$. 68$\%$ and 90$\%$
  confidence level regions are shown.
\label{fig:elliscomp}}}

In order to emulate these results, we perform a similar but Bayesian analysis
with 
the MCMC algorithm: all Standard Model inputs are fixed at their central
empirical values, $\tan \beta=10$ is fixed and $m_0$, $A_0$, $M_{1/2}$ are
  allowed to vary in the MCMC algorithm in order to fit the combined
  posterior probability of dark matter plus other measurements. For this
  comparison, we choose flat priors in $m_0<1$ TeV, $M_{1/2}<1$ TeV and -3
  TeV$<A_0<$3 TeV. The likelihood is calculated as in
  section~\ref{sec:likelihood}. 
The main conclusion from Fig.~\ref{fig:elliscomp} is that the two results 
are similar. If the correct relationship between $\Delta \chi^2$ and
confidence-level were used in Fig.~\ref{fig:elliscomp}a, the confidence level
region could extend out to higher values of $M_{1/2}$. We should note
strictly that, being Bayesian confidence regions as compared to frequentist, 
we do not {\em exactly}\/ compare like with like in
Figs.~\ref{fig:elliscomp}a,b 
but we do expect roughly similar confidence regions in the two cases.
When we perform a similar fit with a larger allowed range of
$m_0<4$ TeV, Fig.~\ref{fig:elliscomp}b deforms due to 
contributions from $h^0$ and fixed-point regions but the preference for
$M_{1/2}<800$ GeV remains. We conclude from this that Ellis {\em et al} 
\/did not scan larger values of $m_0$ where the focus
point regime resides.
The procedure of Ellis {\em et al} \/is not suited
for including the $h^0$ and fixed-point regions, since then there is no unique
solution of $m_0$ which provides the central value of $\Omega_{DM} h^2$. 
If we then additionally
include smearing due to $\tan \beta$ in Fig.~\ref{fig:elliscomp}b with a flat
prior, the $A^0$-pole  
region extends the region of valid $M_{1/2}$ out to higher values $>1$
TeV. Allowing variations of Standard Model input parameters produces further
smearing in the fits until, finally, Fig.~\ref{fig:newprior}a is obtained.

\acknowledgments
This work has been partially supported by STFC\@. We thank R Rattazzi for
discussions which lead to the re-examination of priors. We also thank the
Cambridge SUSY working group and T Plehn for helpful discussions and
observations.  
The computational work has been performed using the Cambridge eScience CAMGRID
computing facility, with the invaluable help of M Calleja.

\end{document}